\documentstyle[axodraw,12pt]{article}

\advance\textheight by \topskip
\newcommand{\sla}{\kern -5.4pt /}
\newcommand{\slaar}{\kern -7. pt / \kern 3.pt}
\newcommand{\Dir}{\kern -6.4pt\Big{/}}%su lettere italiane minuscole
\newcommand{\Dirin}{\kern -10.4pt\Big{/}\kern 4.4pt}
\newcommand{\DDir}{\kern -7.6pt\Big{/}}%su lettere italiane maiuscole
\newcommand{\DGir}{\kern -6.0pt\Big{/}}%su lettere greche

\newcommand{\ra}{\rightarrow}
\newcommand{\be}{\begin{equation}}
\newcommand{\ee}{\end{equation}}
\newcommand{\bea}{\begin{eqnarray}}
\newcommand{\eea}{\end{eqnarray}}
\newcommand{\beanon}{\begin{eqnarray*}}
\newcommand{\eeanon}{\end{eqnarray*}}
\newcommand{\ba}{\begin{array}}
\newcommand{\ea}{\end{array}}
\newcommand{\bd}{\begin{description}}
\newcommand{\ed}{\end{description}}
\newcommand{\bi}{\begin{itemize}}
\newcommand{\ei}{\end{itemize}}
\newcommand{\ben}{\begin{enumerate}}
\newcommand{\een}{\end{enumerate}}
\newcommand{\bc}{\begin{center}}
\newcommand{\ec}{\end{center}}

\newcommand{\ol}{\overline}

\newcommand{\ar}{\rightarrow}

\newcommand{\hsk}{\hskip 10 pt\noindent}

\newcommand{\gtap}{\stackrel{\displaystyle >}{\,_{\! \,_{\displaystyle
\sim}}}}  %maggiore circa uguale
\def\wm{M_{_W}}
\def\zm{M_{_Z}}
\def\gf{G_{\mu}}
\def\pd{{\it production $\times$ decay\ }}
\def\app #1 #2 #3 {{\it  Acta Phys.Polon.} {#1} (#2) #3\ }
\def\ap #1 #2 #3 {{\it Ann. Phys. }{ #1} (#2) #3\ }
\def\intj #1 #2 #3{{\it Int. J. Mod. Phys.} {#1} (#2) #3\ }
\def\hpa #1 #2 #3{{\it Helv. Phys. Acta. }{ #1} #2) #3\ }
\def\mpl #1 #2 #3 {{\it Mod.~Phys.~Lett.} {#1} (#2) #3\ }
\def\np #1 #2 #3 {{\it Nucl.~Phys.} {#1} (#2) #3\ }
\def\pl #1 #2 #3 {{\it Phys.~Lett.} {#1} (#2) #3\ }
\def\pr #1 #2 #3 {{\it Phys.~Rev.} {#1} (#2) #3\ }
\def\prep #1 #2 #3 {{\it Phys.~Rep.} {#1} (#2) #3\ }
\def\prl #1 #2 #3 {{\it Phys.~Rev.~Lett.} {#1} (#2) #3\ }
\def\rmp #1 #2 #3 {{\it Rev. Mod. Phys.} {#1} (#2) #3\ }
\def\zp #1 #2 #3 {{\it Z.~Phys.} {#1} (#2) #3\ }
\def\cpc #1 #2 #3 {{\it Comp. Phys. Commun.} {#1} (#2) #3\ }
\def\xx #1 #2 #3 {{#1}, (#2) #3\ }

\begin{document}
\tolerance=100000
\thispagestyle{empty}
\setcounter{page}{0}

\begin{flushright}
{\large DFTT 74/96}\\
{\rm May 1997\hspace*{.5 truecm}}\\
\end{flushright}

\vspace*{\fill}

{\Large \bf \noindent
Semileptonic six fermion processes at future $e^+ e^-$ colliders: signal
and irreducible background for top and WWZ physics.
\footnote{ Work supported in part by Ministero
dell' Universit\`a e della Ricerca Scientifica.\\[2 mm]
e-mail: accomando,ballestrero,pizzio@to.infn.it}}\\[2.cm]
\bc
{\large Elena Accomando, Alessandro Ballestrero and Marco Pizzio}\\[.3 cm]
{\it I.N.F.N., Sezione di Torino, Italy}\\
{\it and}\\
{\it Dipartimento di Fisica Teorica, Universit\`a di Torino, Italy}\\
{\it v. Giuria 1, 10125 Torino, Italy.}\\
\ec

\vspace*{\fill}

\begin{abstract}
{\normalsize\noindent
 We compute several total and differential cross sections relevant to top
and WWZ physics at future $e^+e^-$ colliders taking into account the full set 
of Feynman diagrams for six fermion final states. We also include in our
calculations initial state radiation and beamstrahlung effects, and the
most important QCD corrections in an approximate  (naive) form.  We compare 
such a complete approach with  \pd approximation 
and we suggest that in many physical studies the former is needed. }
\end{abstract}
.  
\vspace*{\fill}
\newpage

\section{ Introduction }
Since the advent of helicity amplitude methods \cite{ha1} \cite{ha2}
\cite{method} , various processes 
with many  particles and a great number of tree level Feynman diagrams  
have been computed successfully. These processes become more and more
important at high energies for present and future accelerators. In $e^+ e^-$
physics, with the advent of LEP2, WW and Higgs studies are now being performed
with the help of the electroweak dedicated four fermion codes
\cite{fort} \cite{cpc}  \cite{wweg} \cite{dpeg}. 
Most of them compute the amplitudes taking into full account all Feynman
diagrams  and some  can provide cross sections for all possible 4 
fermion processes. 
Previous computations relied on calculating the differential cross sections 
for $e^+e^- \ra W^+W^-$,  $e^+e^- \ra HZ$ or  $e^+e^- \ra ZZ$ and folding them
with the differential branching for the  successive decay of W's, Z's and H.
With such a method it has been possible to compute strong and electroweak
corrections to production and decay. However with such \pd  approximation, 
one is of course 
neglecting the contributions due to many Feynman diagrams as well as
 the off-shellness of signal diagrams. Spin correlation effects are also
disregarded, as one sums over all possible final helicities of  the
intermediate state W's  and Z's, and then computes their decay as 
unpolarized
particles. Only with the complete calculations it is possible to determine the
magnitude of such  approximations.   Various phenomenological analyses 
\cite{ww} \cite{hig} 
 have shown that, even if \pd
may often be reliable for fully extrapolated total cross sections, 
it is certainly not a good approximation when experimental cuts are
implemented and when one is interested in  differential cross sections.

When future $e^+ e^-$ colliders will come into operation, 
states with even more final state particles will become of the utmost interest.
In particular all physics regarding  top studies and 
 three vector boson production  will be concerned with six fermion final 
states. For such a reason we  compute in this paper all semileptonic
six fermion processes with four final quarks and compare in detail the full 
calculation versus
\pd approximations for some cross sections
and distributions of interest. We consider for the time being only the case 
in which one has no Higgs which can decay into two W's.  Of course if 
that would not be the case, Higgs production itself would end up in a 
six fermion process. 
 Top  and Higgs physics have already been
analyzed in this respect within a four particles ($b \bar b W^+W^-$) final 
state  approximation \cite{bbww}.
Other groups are at present working on six fermion final state physics 
\cite{pv}\cite{kuri}. In particular in ref.\cite{pv} the processes
$e^+e^-\ar \mu^+\mu^-\tau^- \bar\nu_\tau u\bar d$ and 
$e^+e^-\ar \mu^+\mu^- e^- \bar\nu_e u\bar d$ have been computed 
with the inclusion of an intermediate Higgs contributions and  these reactions
have been used to discuss  the relevance
of six fermion full calculations for Higgs physics.
Other computations \cite{kuri} have been performed on 
$e^+e^-\ar b \bar b u\bar d \mu^- \bar \nu_\mu$ which is instead of major
interest to top physics in the continuum.

The plan of the paper is the following: in section~2 we analyze some general
properties of all possible
six fermion processes for $e^+ e^-$ colliders, in section~3 we give some 
details of the calculation,
in section~4 we present our results for top physics, while section~5 will
cover WWZ physics. Some considerations and outlooks are given in the
conclusions.

\section {$\bf e^+ e^- \ra six fermion$ processes}
 
All processes $e^+ e^- \ra six fermions$ can be divided
according to their final states. As for four fermion final states, we can 
identify Charged Currents (CC), Neutral Currents (NC) and 
Mixed processes (MIX) \cite {class}\cite{wweg} \cite{cpc}. 
The final states for six fermion 
processes can in fact be deduced from the corresponding ones for four fermions,
 adding to them a particle antiparticle pair. 

For four fermions the convention is to call
CC those processes in which the final particles can form two W's and not
two Z's (e. g. $\mu \bar \nu u \bar d $),
NC those in which they form two Z's, but cannot form
two W's   (e. g. $u \bar u \mu^+  \mu^-$) and MIX those in which
they can form both (e. g. $u \bar u d \bar d$). 

Of course this implies that in every single diagram
of CC processes there is at least one charged current, but it does not exclude
 neutral currents. Also for NC one can  have charged 
currents (as it is the case for $\nu_e \bar \nu_e b \bar b$) together with 
neutral ones. 

We will from now on call CC the six fermion processes in which
the final particles  can form two W's and a Z but not three Z's (e. g. $\mu
\bar \nu u \bar d e^+ e^- $),
NC those in which three Z's and not two W's and a Z can be formed 
(e. g. $u \bar u \mu^+  \mu^- e^+ e^-$) , MIX
those in which  both can be formed (e. g. $u \bar u d \bar d e^+ e^-$).

Six fermion CC processes are of particular interest because they are related
to top, WWZ and Higgs physics. If we consider for instance a final state
like $b \bar b \mu \bar \nu_\mu u \bar d$, this may be produced  by the decay
of WWZ or come from two W's and two b's, which in turn may descend from
two tops. WWZ intermediate state may be due to Higgs Z production with
an Higgs decaying into two W's.  Final states like $\nu_e \bar \nu_e \mu \bar 
\nu_\mu u\bar d$ may be produced by the WW $\ra$ H fusion diagram (which 
dominates upon Higgs strahlung at energies $\gtap$ 500 GeV) and successive
decay of the Higgs as above. Of course the various channels are not separated
in the reality, and many more diagrams (irreducible background) contribute to
such processes. They can be distinguished only for their different contribution
to various zones of the phase space and can therefore be disentangled by 
applying
experimental cuts. In order to study such cuts, and to evaluate the magnitude
of the various contributions after they have been applied, it is fundamental
 to use the full calculation.
This is precisely what we  do for  CC semileptonic 
processes, i. e. for the case in which  two of the particles which can 
reconstruct a W are a lepton and a neutrino. With such a prescription,
we are excluding all processes with four quarks reconstructing two W's, as
for instance $\mu^+ \mu^-  u \bar d \bar c s$ or the fully hadronic
$b \bar b u \bar d \bar c s$. It would have been straightforward to 
include in our analysis such processes, but it is meaningless to consider them
without  NC and MIX events. When four quarks are present in the final state,
different flavors cannot be experimentally recognized  (apart from b-tagging),
so that $\mu^+ \mu^- u \bar d \bar c s$ (CC) cannot be distinguished from
$\mu^+ \mu^- u \bar d \bar u d$ (MIX) or $\mu^+ \mu^- u \bar u s \bar s$ (NC).
This consideration shows that on one side considering non semileptonic
processes makes computations  much longer if not harder, on the other that
the semileptonic events will be much cleaner and easy to study also
from an experimental point of view.

Keeping in mind that for the hadronic case all CC, NC, MIX processes have to
be considered, let us briefly discuss  pure NC events.
In principle they are of course relevant as they can be produced by three
Z's or from two Z's and a separate pair. So it seems that their main interest
within the Standard Model corresponds to Higgs strahlung or ZZ fusion with 
successive Higgs decay in two Z's. These events are  depressed
with respect to the corresponding Higgs $\ra$ WW ones by the the different
branching ratios and for the higher cross sections of $e^+e^-\ra H \nu_e
\bar \nu_e$ with respect to $e^+e^-\ra H e^+e^-$. Also for this kind of 
signals, if the
final state corresponds to two Z's decaying purely hadronically, one is
in principle forced to consider CC and MIX processes together with them.  

In the following we will consider only the semileptonic six fermion 
CC processes for the case in which there is no Higgs contribution. The full
calculation will be used to discuss top top events in the continuum and
WWZ events. 

For what concerns top physics, after its discovery  and the
 measurement of its mass (m$_t$= 174 $\pm$ 6 GeV) \cite{top}, it is extremely
important to determine its properties with high-precision.  Just because of
the very large mass, they are most likely affected by the Higgs particles,
and  the top will surely play a crucial role in any theory of flavour
dynamics. An extremely important characteristic of the top is  its
lifetime, which is much shorter than the time scale of strong interactions,
allowing to study it in the context of perturbative QCD. Even if a proper
toponium resonance cannot be formed at such a high top mass, perturbative
effects result in a sharp rise of the cross section at threshold \cite{kuhn},
superimposed on the natural sharp rise due to the opening of a new channel.
For such a reason, in order to measure
top  mass the best strategy is probably to have the future $e^+e^-$ collider
run at $t\bar t$ threshold, while its static properties, such as magnetic and
electric dipole moments \cite{ttform} will be measured with high accuracy in 
the continuum,
at higher energies. Our  results refer only to the continuum top top
production. Our tree level matrix elements for the complete calculation of the
various final states, cannot in fact take into account the
above mentioned threshold corrections. They are nevertheless useful to
understand at any energy the relevance of the \pd
approximation  on the various physical quantities to be measured.
Also at threshold they can for instance be used to estimate the irreducible
background due to all non double resonant diagrams.

Another reason why top top in the continuum has to be carefully computed 
 is that it constitutes
a severe background to WWZ events. These events are important on their own,
as they allow  to measure  gauge couplings. In particular 
the quartic gauge coupling and its possible deviations from SM will be studied
in this reaction at future $e^+e^-$ colliders. Several studies \cite{wwz,miya}
have already analyzed three vector boson production and anomalous gauge
couplings. Actually LHC measurements on VV+X (V=W,Z) final states will reach a
better quartic gauge sensitivity than the WWZ measurement at 500 GeV $e^+e^-$
colliders.\cite{doba,miya} Nevertheless the cleaner $e^+e^-$ experimental
environment as well as the possibility of achieving higher energies or
luminosities require a careful theoretical study.
We therefore  consider all possible contributions to WWZ signal and
backgrounds as they can be defined in the realistic six fermion final state.
In particular we analyze the sum of all semileptonic channels with four quarks
in which  WWZ can decay. We sum over all final states not containing
$b$ quarks, so that b-tagging can help to reduce most of $t \bar t$ background.

\section {Six fermion processes calculation}\label{sixf}
Among the processes that our program SIXPHACT computes, 
one can distinguish CC209, CC418, CC836.  As customary  in four fermion physics,
the numbers after CC refer to the Feynman diagrams of the process. 
Those we have indicated count  diagrams without Higgs only. 
 Examples of CC209 are $e^+e^- \ra \mu \bar \nu_\mu u \bar d b \bar b$
and $e^+e^- \ra \mu \bar \nu_\mu u \bar d s \bar s$. The number of Higgs
diagrams for these processes is respectively 23 and 1.
For processes with b's in the final state, we exactly account for the masses
both in the matrix elements and in the phase space. 
Apart from the masses and couplings, all processes of this kind have the same
matrix elements. The CC418 processes can be obtained from the previous ones
with the exchange of two identical fermions in the final state or with
the exchange of a fermion from the initial state with one from the final.
So we can have two different types: $\rm CC418_f$ for processes like
$e^+e^- \ra \mu \bar \nu_\mu u \bar d u \bar u$ or $e^+e^- \ra \mu \bar \nu_\mu
u \bar d d\bar d$ and $\rm CC418_i$
for the ones of the type $e^+e^- \ra e^- \bar \nu_e u \bar d s \bar s$
 Analogously, CC836 are obtained by CC209 with two exchanges,
 one between   initial and  final and the other between two final particles, as
 it happens for $e^+e^- \ra e^- \bar \nu_e u \bar d u \bar u$. 
There are  processes which can have  more exchanges, like
$e^+e^- \ra e^- \bar \nu_e u \bar d e^-e^+$ (CC1248).
These are however of no interest for top physics and have a very low
cross section as compared to the sum of four quarks CC final states, 
and  we will  not consider them.

The complexity of the problem is not only due to the great number of  rather
complex diagrams: the integration on the  phase space 
 is at least thirteen dimensional, but it  becomes  seventeen dimensional 
for the more realistic case in which  initial state radiation (ISR) and 
beamstrahlung (BST) are accounted for. For such reasons it is extremely 
important to use a method for computing helicity 
amplitudes which allows a very fast and precise computation.  
We have used to this end  PHACT \cite{phact}, a set of routines based on
the method of ref.\cite{method}, in order to generate the fortran code
for the amplitudes. The resulting program SIXPHACT is able to 
compute the cross sections in reasonable
cpu time (of the order of a few hours on a alphastation) with such a precision as to allow
at the same time to generate differential cross sections (distributions) 
without visible statistical fluctuations. This normally requires a precision of
the order of at least
.5\% on every bin and an overall precision at least ten times better.

As far as the matrix elements are concerned, 
we have computed all Feynman diagrams by calculating parts of diagrams of
increasing complexity. The method used is particularly suited for this 
procedure and we will briefly explain it.

We start with the subdiagrams corresponding
to a $\gamma$, a $Z$ or a $W$ into a pair of external fermions:

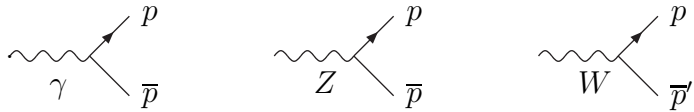
\begin{picture}(420,60)(0,0)
\thicklines

%4
\SetOffset(100,30)
% Zeta ff secondario
\Text(0,0)[]{$.$}
\Photon(0,0)(30,0){2}{3}
\Text(15,-14)[lb]{$\gamma$}
\ArrowLine(30,0)(45,15)
\Text(50,15)[l]{$p$}
\Line (30,0)(45,-15)
\Text(50,-15)[l]{$\ol p$}

\SetOffset(200,30)
% Zeta ff secondario
\Photon(0,0)(30,0){2}{3}
\Text(15,-14)[lb]{$Z$}
\ArrowLine(30,0)(45,15)
\Text(50,15)[l]{$p$}
\Line (30,0)(45,-15)
\Text(50,-15)[l]{$\ol p$}

\SetOffset(300,30)
% Zeta ff secondario
\Photon(0,0)(30,0){2}{3}
\Text(15,-14)[lb]{$W$}
\ArrowLine(30,0)(45,15)
\Text(50,15)[l]{$p$}
\Line (30,0)(45,-15)
\Text(50,-15)[l]{$\ol p'$}
\end {picture}

\noindent 
and then we compute the sum of subdiagrams defined in eqs.~1-8.
Here and in the following, we  use the prime to indicate a particle 
associated in an weak isospin doublet. So if, for instance,  $p$ will 
represent an  up quark, $\ol p'$ will indicate an anti-down quark.
We will also use the symbols $p$, $q$, $P$ to indicate external 
momenta, while $r$, $s$ will correspond to momenta which might be both 
external or internal.  
Eq.~(1) represents the sum of  $\gamma$ and $Z$ insertions
 of the subdiagrams  above in a fermion line:
%eq 1
\be
\vcenter{\hbox{
\begin{picture}(80,60)(-15,-30)
%6
%linea verticale con eventuali impulsi
\ArrowLine(0,-20)(0,20)
\Text(-5,20)[rb]{$r$}
\Text(-5,-20)[rt]{$s$}
% gamma,Zeta ff secondario
\Photon(0,0)(30,0){2}{3}
\Text(10,-15)[lb]{$\gamma Z$}
\ArrowLine(30,0)(45,15)
\Text(50,15)[l]{$p$}
\Line (30,0)(45,-15)
\Text(50,-15)[l]{$\ol p$}
\end{picture}
}}
=
\vcenter{\hbox{
\begin{picture}(80,60)(-15,-30)
%6
%linea verticale con eventuali impulsi
\ArrowLine(0,-20)(0,20)
\Text(-5,20)[rb]{$r$}
\Text(-5,-20)[rt]{$s$}
% gamma,Zeta ff secondario
\Photon(0,0)(30,0){2}{3}
\Text(13,-15)[lb]{$\gamma $}
\ArrowLine(30,0)(45,15)
\Text(50,15)[l]{$p$}
\Line (30,0)(45,-15)
\Text(50,-15)[l]{$\ol p$}
\end{picture}
}}
+
\vcenter{\hbox{
\begin{picture}(80,60)(-15,-30)
%6
\SetOffset(0,0)
%linea verticale con eventuali impulsi
\ArrowLine(0,-20)(0,20)
\Text(-5,20)[rb]{$r$}
\Text(-5,-20)[rt]{$s$}
% gamma,Zeta ff secondario
\Photon(0,0)(30,0){2}{3}
\Text(13,-15)[lb]{$Z$}
\ArrowLine(30,0)(45,15)
\Text(50,15)[l]{$p$}
\Line (30,0)(45,-15)
\Text(50,-15)[l]{$\ol p$}
\end{picture}
}}
\ee

The diagrams of eq.~(2) correspond to a virtual $W$  
"decaying" into four outgoing  fermions:

%eq 2
\bea
\vcenter{\hbox{
\begin{picture}(80,60)(-5,-30)
\Photon(0,0)(30,0){2}{3}
\Text(10,-14)[lb]{$W$}
\Oval(45,0)(18,15)(0)
\Text(37,12)[lt]{$p$ $\ol p'$}
\Text(37,-12)[lb]{$q$ $\ol q$}
\end{picture}
}}
&=&
\vcenter{\hbox{
\begin{picture}(110,90)(-5,-45)
%8
% gamma ff primario piccolo
\Photon(0,0)(30,0){2}{3}
\Text(10,-14)[lb]{$W$}
\ArrowLine(30,0)(48,30)
\Text(40,40)[l]{$p$}
\Line (30,0)(48,-30)
\Text(40,-40)[l]{$\ol p'$}
% gamma,Zeta ff secondario
\Photon(39,15)(69,15){2}{3}
\Text(49,0)[lb]{$\gamma Z$}
\ArrowLine(69,15)(84,30)
\Text(89,30)[l]{$q$}
\Line (69,15)(84,0)
\Text(89,0)[l]{$\ol q$}
\end{picture}
}}
+
\vcenter{\hbox{
\begin{picture}(110,90)(-5,-45)
%9
% gamma ff primario piccolo
\Photon(0,0)(30,0){2}{3}
\Text(10,-14)[lb]{$W$}
\ArrowLine(30,0)(48,30)
\Text(40,40)[l]{$p$}
\Line (30,0)(48,-30)
\Text(40,-40)[l]{$\ol p'$}
% gamma,Zeta ff secondario
\Photon(39,-15)(69,-15){2}{3}
\Text(49,-30)[lb]{$\gamma Z$}
\ArrowLine(69,-15)(84,0)
\Text(89,0)[l]{$q$}
\Line (69,-15)(84,-30)
\Text(89,-30)[l]{$\ol q$}
\end{picture}
}}\nonumber \\
&&+
\vcenter{\hbox{
\begin{picture}(110,90)(-5,-45)
%8
% gamma ff primario piccolo
\Photon(0,0)(30,0){2}{3}
\Text(10,-14)[lb]{$W$}
\ArrowLine(30,0)(48,30)
\Text(40,40)[l]{$q$}
\Line (30,0)(48,-30)
\Text(40,-40)[l]{$\ol q$}
% gamma,Zeta ff secondario
\Photon(39,15)(69,15){2}{3}
\Text(49,0)[lb]{$W$}
\ArrowLine(69,15)(84,30)
\Text(89,30)[l]{$p$}
\Line (69,15)(84,0)
\Text(89,0)[l]{$\ol p'$}
\end{picture}
}}
+
\vcenter{\hbox{
\begin{picture}(110,90)(-5,-45)
%10
\Photon(0,0)(30,0){2}{3}
\Text(10,-14)[lb]{$W$}
\Photon(30,0)(45,25){2}{3}
\Photon(30,0)(45,-25){2}{3}
\ArrowLine(45,-25)(60,-10)
\Text(65,-10)[l]{$q$}
\Line (45,-25)(60,-40)
\Text(65,-40)[l]{$\ol q$}
\ArrowLine(45,25)(60,40)
\Text(65,40)[l]{$p$}
\Line (45,25)(60,10)
\Text(65,10)[l]{$\ol p'$}
\end{picture}
}}
\eea
\noindent Notice that the third subdiagram refers to the cases of an external
incoming $W^-$($W^+$) and up(down) type flavour for $q\bar q$. In  the other
cases, the internal $W$ is not attached to $q$ but to $\bar q$.

We do not compute 
the sums of diagrams of eq.~(2) when $q$ and $\ol q$ correspond to incoming
$e^- e^+$. 
When a whole diagram is divided by a W propagator in two parts
containing four fermions each, one may choose
 to compute one of the two as a subdiagram (2) and
our choice is to compute the part not  containing  $e^- e^+$.

The diagrams of eq.~(3) correspond 
to a virtual $\gamma$ (or $Z$) "decaying" into four
outgoing  fermions:

%eq 3
\bea
%13
\vcenter{\hbox{
\begin{picture}(70,60)(-5,-30)
%13
\Photon(0,0)(30,0){2}{3}
\Text(2,-16)[lb]{$\gamma (Z)$}
\Boxc(42,0)(24,30)
\Text(34,12)[lt]{$p$ $\ol p'$}
\Text(34,-12)[lb]{$q$ $\ol q'$}
\end{picture}
}}
%&=&
=
\vcenter{\hbox{
\begin{picture}(110,90)(-5,-45)
%8
% gamma ff primario piccolo
\Photon(0,0)(30,0){2}{3}
\Text(2,-16)[lb]{$\gamma (Z)$}
\ArrowLine(30,0)(48,30)
\Text(40,40)[l]{$p$}
\Line (30,0)(48,-30)
\Text(40,-40)[l]{$\ol p'$}
% gamma,Zeta ff secondario
\Photon(39,15)(69,15){2}{3}
\Text(49,0)[lb]{$W$}
\ArrowLine(69,15)(84,30)
\Text(89,30)[l]{$q$}
\Line (69,15)(84,0)
\Text(89,0)[l]{$\ol q'$}
\end{picture}
}}
+
\vcenter{\hbox{
\begin{picture}(110,90)(-5,-45)
%9
% gamma ff primario piccolo
\Photon(0,0)(30,0){2}{3}
\Text(2,-16)[lb]{$\gamma (Z)$}
\ArrowLine(30,0)(48,30)
\Text(40,40)[l]{$p$}
\Line (30,0)(48,-30)
\Text(40,-40)[l]{$\ol p'$}
% gamma,Zeta ff secondario
\Photon(39,-15)(69,-15){2}{3}
\Text(54,-28)[lb]{$W$}
\ArrowLine(69,-15)(84,0)
\Text(89,0)[l]{$q$}
\Line (69,-15)(84,-30)
\Text(89,-30)[l]{$\ol q'$}
\end{picture}
}}+\hsk\hsk\hsk\nonumber \\
\hsk\hsk\hsk
\vcenter{\hbox{
\begin{picture}(110,90)(-5,-45)
%8
% gamma ff primario piccolo
\Photon(0,0)(30,0){2}{3}
\Text(2,-16)[lb]{$\gamma (Z)$}
\ArrowLine(30,0)(48,30)
\Text(40,40)[l]{$q$}
\Line (30,0)(48,-30)
\Text(40,-40)[l]{$\ol q'$}
% gamma,Zeta ff secondario
\Photon(39,15)(69,15){2}{3}
\Text(49,0)[lb]{$W$}
\ArrowLine(69,15)(84,30)
\Text(89,30)[l]{$p$}
\Line (69,15)(84,0)
\Text(89,0)[l]{$\ol p'$}
\end{picture}
}}
+
\vcenter{\hbox{
\begin{picture}(110,90)(-5,-45)
%9
% gamma ff primario piccolo
\Photon(0,0)(30,0){2}{3}
\Text(2,-16)[lb]{$\gamma (Z)$}
\ArrowLine(30,0)(48,30)
\Text(40,40)[l]{$q$}
\Line (30,0)(48,-30)
\Text(40,-40)[l]{$\ol q'$}
% gamma,Zeta ff secondario
\Photon(39,-15)(69,-15){2}{3}
\Text(54,-28)[lb]{$W$}
\ArrowLine(69,-15)(84,0)
\Text(89,0)[l]{$p$}
\Line (69,-15)(84,-30)
\Text(89,-30)[l]{$\ol p'$}
\end{picture}
}}
+
\vcenter{\hbox{
\begin{picture}(110,90)(-5,-45)
%10
\Photon(0,0)(30,0){2}{3}
\Text(2,-16)[lb]{$\gamma (Z)$}
\Photon(30,0)(45,25){2}{3}
\Photon(30,0)(45,-25){2}{3}
\ArrowLine(45,-25)(60,-10)
\Text(65,-10)[l]{$q$}
\Line (45,-25)(60,-40)
\Text(65,-40)[l]{$\ol q'$}
\ArrowLine(45,25)(60,40)
\Text(65,40)[l]{$p$}
\Line (45,25)(60,10)
\Text(65,10)[l]{$\ol p'$}
\end{picture}
}}
\eea
  The diagrams of
Eq.~(4) represent the  sum of  $\gamma$ and $Z$ insertions
 in a fermion line of the subdiagrams defined in eq.~(3):
%eq 4
\be
\vcenter{\hbox{
\begin{picture}(80,60)(-10,-30)
%15
%linea verticale con eventuali impulsi
\ArrowLine(0,-20)(0,20)
\Text(-5,20)[rb]{$r$}
\Text(-5,-20)[rt]{$s$}
% gamma ff primario piccolo
\Photon(0,0)(30,0){2}{3}
\Text(8,-14)[lb]{$\gamma Z$}
\Boxc(42,0)(24,30)
\Text(34,12)[lt]{$p$ $\ol p'$}
\Text(34,-12)[lb]{$q$ $\ol q'$}
\end{picture}
}}
=
\vcenter{\hbox{
\begin{picture}(80,60)(-10,-30)
%15
%linea verticale con eventuali impulsi
\ArrowLine(0,-20)(0,20)
\Text(-5,20)[rb]{$r$}
\Text(-5,-20)[rt]{$s$}
% gamma ff primario piccolo
\Photon(0,0)(30,0){2}{3}
\Text(10,-14)[lb]{$\gamma$}
\Boxc(42,0)(24,30)
\Text(34,12)[lt]{$p$ $\ol p'$}
\Text(34,-12)[lb]{$q$ $\ol q'$}
\end{picture}
}}
+
\vcenter{\hbox{
\begin{picture}(80,60)(-10,-30)
%15
%linea verticale con eventuali impulsi
\ArrowLine(0,-20)(0,20)
\Text(-5,20)[rb]{$r$}
\Text(-5,-20)[rt]{$s$}
% gamma ff primario piccolo
\Photon(0,0)(30,0){2}{3}
\Text(10,-14)[lb]{$Z$}
\Boxc(42,0)(24,30)
\Text(34,12)[lt]{$p$ $\ol  p'$}
\Text(34,-12)[lb]{$q$ $\ol q'$}
\end{picture}
}}
\ee
The remaining subdiagrams to compute correspond  to the emission of 4 fermions
from the upper part of a fermion
line.  Let's start with the case in which the four fermions are all outgoing. 
%eq 5
\be
\vcenter{\hbox{
\begin{picture}(40,110)(-10,-55)
%12
\Boxc(0,0)(24,30)
\Text(-8,12)[lt]{$p$ $\ol p'$}
\Text(-8,-12)[lb]{$q$ $\ol q$}
\Line (0,15)(0,34)
\Line (0,-15)(0,-34)
\Text(-3,34)[rb]{$P$}
%\Text(-3,-34)[rt]{$1$}
\end{picture}
}}
=
\vcenter{\hbox{
\begin{picture}(80,110)(-10,-55)
%11
%linea verticale con eventuali impulsi
\ArrowLine(0,-45)(0,45)
\Text(-5,45)[rb]{$P$}
%\Text(-5,-45)[rt]{$1$}
% gamma,Zeta ff secondario
\Photon(0,25)(30,25){2}{3}
\Text(12,10)[lb]{$W$}
\ArrowLine(30,25)(45,40)
\Text(50,40)[l]{$p$}
\Line (30,25)(45,10)
\Text(50,10)[l]{$\ol p'$}
% gamma,Zeta ff secondario
\Photon(0,-25)(30,-25){2}{3}
\Text(10,-40)[lb]{$\gamma Z$}
\ArrowLine(30,-25)(45,-10)
\Text(50,-10)[l]{$q$}
\Line (30,-25)(45,-40)
\Text(50,-40)[l]{$\ol q$}
\end{picture}
}}
+
\vcenter{\hbox{
\begin{picture}(80,110)(-10,-55)
%11
%linea verticale con eventuali impulsi
\ArrowLine(0,-45)(0,45)
\Text(-5,45)[rb]{$P$}
%\Text(-5,-45)[rt]{$1$}
% gamma,Zeta ff secondario
\Photon(0,25)(30,25){2}{3}
\Text(10,10)[lb]{$\gamma Z$}
\ArrowLine(30,25)(45,40)
\Text(50,40)[l]{$q$}
\Line (30,25)(45,10)
\Text(50,10)[l]{$\ol q$}
% gamma,Zeta ff secondario
\Photon(0,-25)(30,-25){2}{3}
\Text(12,-40)[lb]{$W$}
\ArrowLine(30,-25)(45,-10)
\Text(50,-10)[l]{$p$}
\Line (30,-25)(45,-40)
\Text(50,-40)[l]{$\ol p'$}
\end{picture}
}}
+
\vcenter{\hbox{
\begin{picture}(80,110)(-10,-55)
%17
%linea verticale con eventuali impulsi
\ArrowLine(0,-20)(0,20)
\Text(-5,20)[rb]{$P$}
%\Text(-5,-20)[rt]{$1$}
% gamma ff primario piccolo
\Photon(0,0)(30,0){2}{3}
\Text(10,-14)[lb]{$W$}
\Oval(45,0)(18,15)(0)
\Text(37,12)[lt]{$p$ $\ol p'$}
\Text(37,-12)[lb]{$q$ $\ol q$}
\end{picture}
}}
\ee

%eq 6
\be
\vcenter{\hbox{
\begin{picture}(40,110)(-10,-55)
%11
\Boxc(0,0)(24,30)
\Text(-8,12)[lt]{$p$ $\ol p'$}
\Text(-8,-12)[lb]{$q$ $\ol q'$}
\Line (0,15)(0,34)
\Line (0,-15)(0,-34)
\Text(-3,34)[rb]{$P$}
%\Text(-3,-34)[rt]{$1$}
\end{picture}
}}
=
\vcenter{\hbox{
\begin{picture}(80,110)(-10,-55)
%11
%linea verticale con eventuali impulsi
\ArrowLine(0,-45)(0,45)
\Text(-5,45)[rb]{$P$}
%\Text(-5,-45)[rt]{$1$}
% gamma,Zeta ff secondario
\Photon(0,25)(30,25){2}{3}
\Text(12,10)[lb]{$W$}
\ArrowLine(30,25)(45,40)
\Text(50,40)[l]{$p$}
\Line (30,25)(45,10)
\Text(50,10)[l]{$\ol p'$}
% gamma,Zeta ff secondario
\Photon(0,-25)(30,-25){2}{3}
\Text(12,-40)[lb]{$W$}
\ArrowLine(30,-25)(45,-10)
\Text(50,-10)[l]{$q$}
\Line (30,-25)(45,-40)
\Text(50,-40)[l]{$\ol q'$}
\end{picture}
}}
+
\vcenter{\hbox{
\begin{picture}(80,110)(-10,-55)
%16
%linea verticale con eventuali impulsi
\ArrowLine(0,-20)(0,20)
\Text(-5,20)[rb]{$P$}
%\Text(-5,-20)[rt]{$1$}
% gamma ff primario piccolo
\Photon(0,0)(30,0){2}{3}
\Text(8,-14)[lb]{$\gamma Z$}
\Boxc(42,0)(24,30)
\Text(34,12)[lt]{$p$ $\ol p'$}
\Text(34,-12)[lb]{$q$ $\ol q'$}
\end{picture}
}}
\ee

When two of the emitted fermions are the incoming $e^+$ $e^-$, we use a
different symbol and a somewhat different definition as the term with
the "decay" of a virtual $W$, $Z$  or $\gamma$ into four fermions is missing:

%eq 7
\be
\vcenter{\hbox{
\begin{picture}(40,110)(-10,-55)
%13
\Oval(0,0)(18,15)(0)
\Text(-8,12)[lt]{$p$ $\ol p'$}
\Text(-11,-12)[lb]{$e^- e^+$}
\Line (0,18)(0,34)
\Line (0,-18)(0,-34)
\Text(-3,34)[rb]{$P$}
%\Text(-3,-34)[rt]{$1$}
\end{picture}
}}
=
\vcenter{\hbox{
\begin{picture}(80,110)(-10,-55)
%11
%linea verticale con eventuali impulsi
\ArrowLine(0,-45)(0,45)
\Text(-5,45)[rb]{$P$}
%\Text(-5,-45)[rt]{$1$}
% gamma,Zeta ff secondario
\Photon(0,25)(30,25){2}{3}
\Text(12,10)[lb]{$W$}
\ArrowLine(30,25)(45,40)
\Text(50,40)[l]{$p$}
\Line (30,25)(45,10)
\Text(50,10)[l]{$\ol p'$}
% gamma,Zeta ff secondario
\Photon(0,-25)(30,-25){2}{3}
\Text(10,-40)[lb]{$\gamma Z$}
\ArrowLine(30,-25)(45,-10)
\Text(50,-10)[l]{$e^+$}
\Line (30,-25)(45,-40)
\Text(50,-40)[l]{$e^-$}
\end{picture}
}}
+
\vcenter{\hbox{
\begin{picture}(80,110)(-10,-55)
%11
%linea verticale con eventuali impulsi
\ArrowLine(0,-45)(0,45)
\Text(-5,45)[rb]{$P$}
%\Text(-5,-45)[rt]{$1$}
% gamma,Zeta ff secondario
\Photon(0,25)(30,25){2}{3}
\Text(10,10)[lb]{$\gamma Z$}
\ArrowLine(30,25)(45,40)
\Text(50,40)[l]{$e^+$}
\Line (30,25)(45,10)
\Text(50,10)[l]{$e^-$}
% gamma,Zeta ff secondario
\Photon(0,-25)(30,-25){2}{3}
\Text(12,-40)[lb]{$W$}
\ArrowLine(30,-25)(45,-10)
\Text(50,-10)[l]{$p$}
\Line (30,-25)(45,-40)
\Text(50,-40)[l]{$\ol p'$}
\end{picture}
}}
\ee

%eq 8
\be
\vcenter{\hbox{
\begin{picture}(40,110)(-10,-55)
%13
\Oval(0,0)(18,15)(0)
\Text(-8,12)[lt]{$q$ $\ol q$}
\Text(-11,-12)[lb]{$e^- e^+$}
\Line (0,18)(0,34)
\Line (0,-18)(0,-34)
\Text(-3,34)[rb]{$P$}
%\Text(-3,-34)[rt]{$1$}
\end{picture}
}}
=
\vcenter{\hbox{
\begin{picture}(80,110)(-10,-55)
%11
%linea verticale con eventuali impulsi
\ArrowLine(0,-45)(0,45)
\Text(-5,45)[rb]{$P$}
%\Text(-5,-45)[rt]{$1$}
% gamma,Zeta ff secondario
\Photon(0,25)(30,25){2}{3}
\Text(10,10)[lb]{$\gamma Z$}
\ArrowLine(30,25)(45,40)
\Text(50,40)[l]{$q$}
\Line (30,25)(45,10)
\Text(50,10)[l]{$\ol q$}
% gamma,Zeta ff secondario
\Photon(0,-25)(30,-25){2}{3}
\Text(10,-40)[lb]{$\gamma Z$}
\ArrowLine(30,-25)(45,-10)
\Text(50,-10)[l]{$e^+$}
\Line (30,-25)(45,-40)
\Text(50,-40)[l]{$e^-$}
\end{picture}
}}
+
\vcenter{\hbox{
\begin{picture}(80,110)(-10,-55)
%11
%linea verticale con eventuali impulsi
\ArrowLine(0,-45)(0,45)
\Text(-5,45)[rb]{$P$}
%\Text(-5,-45)[rt]{$1$}
% gamma,Zeta ff secondario
\Photon(0,25)(30,25){2}{3}
\Text(10,10)[lb]{$\gamma Z$}
\ArrowLine(30,25)(45,40)
\Text(50,40)[l]{$e^+$}
\Line (30,25)(45,10)
\Text(50,10)[l]{$e^-$}
% gamma,Zeta ff secondario
\Photon(0,-25)(30,-25){2}{3}
\Text(10,-40)[lb]{$\gamma Z$}
\ArrowLine(30,-25)(45,-10)
\Text(50,-10)[l]{$q$}
\Line (30,-25)(45,-40)
\Text(50,-40)[l]{$\ol q$}
\end{picture}
}}
\ee

With the previous definitions, having computed the appropriate subdiagrams,
the 209 diagrams corresponding to the process $e^+e^- \ra u\ \bar d\  \mu\ \bar\nu_\mu\
  b\ \bar b$ reduce to the ones represented in fig.~1.
The other amplitudes for CC418 and CC836 are then obtained from these ones
just exchanging the momenta of the appropriate particles.

The phase space integration of these amplitudes requires particular care
for different  resonant structures of the various sets of diagrams.
From a kinematical point of view a six particles phase space integration
might be thought as a sequence of two-body  decays into intermediate states
 that can be formed with whatever combination of external momenta. This
sequence ends up into the physical outgoing particles. A set of integration
variables is constituted by the invariant masses of the intermediate 
states  and by the angular coordinates  of the decay
products in their rest frame. With a proper change of variable on some
invariant masses we have smoothed out the resonant Breit-Wigner peaks.
Different combinations of momenta have been evaluated for each 
subset of diagrams, in particular for what concerns the background
 contributions.
We have essentially used three parametrizations. For the integration of
diagrams of top signal we have used a  $t \bar tW^+W^-$ resonant phase space.
In it, we choose as integration variables the masses of the two tops and of the
two W's, on which we perform the above mentioned change of variables.
Analogously we have used a WWZ resonant phase space for off shell production
of three vector bosons and a WW resonant one for top  background.
The numerical integration has been performed with Vegas \cite{vegas}.

The computations we present take into account both initial state radiation
 and beamstrahlung. ISR is introduced via the structure function method
\cite{sf}.  
The integration with ISR is 15-dimensional. BST
 is introduced with a link to the program 
CIRCE~\cite{circe}. This program generates distributions of the
fractions  $x_1,x_2$ of momenta  carried by the incoming electron and positron
 after BST. Either one integrates over these distribution 
 functions or one makes use of the unweighted  generation by CIRCE of the 
above fractions. We have chosen this second possibility: for every 
integration point we generate with CIRCE a couple  of $x_1,x_2$ and reconstruct
the kinematics accordingly.

As far as QCD corrections are concerned, we have introduced them in the so
called naive QCD approach (NQCD). This amounts to consider that we have
diagrams with vertices which in narrow width approximation (NWA) correspond to 
\pd of W's, Z's, tops. In a fully extrapolated
setup the corresponding correction for the decay  is factorized and amounts
to multiplying the W decay vertex by $ (1+\alpha(m_W)/\pi)^{1/2}$, the
Z one by $(1+\alpha(m_Z)/\pi)^{1/2}$ and the top one by 
\be
\left [ 1-{2\alpha_s(m_t)\over{3\pi}}f(y)\right ]^{1\over 2},
\ \ \ \ \ {\rm with } \ \ y=\left( {m_W\over{m_t}}\right) ^2.
\ee
The function f(x) has been calculated in ref.\cite{twidth} 
and, in the  $m_b/m_t$, $\Gamma_W/M_W\ \ar 0$ limit,
 it can be written in the compact form:
\[
f(y)={2\pi^2\over 3}-{5\over 2}+2\; ln(y)\; ln(1-y)+4Li_2(y)-2y+
\]
\be
{1\over{1+2y}}\left [ (5+4y)\; ln(1-y)+{2y\; ln(y)\over{1-y}}-{4y^3(1-y+ln(y))
\over{(1-y)^2}}\right ]
\ee
As a result of the strong correction, the top width lowers by a factor of 
order 10$\%$. 

We have also included the QCD corrections in the 
naive formulation to the $t\bar t$ production vertices. The first order QCD 
corrections~\cite{Zttcor} to the on-shell top top production total 
cross section can be written as
\begin{eqnarray}
\sigma =  \sigma_{VV}\left ( 1+{4\alpha_s(m_t)\over{3\pi}}K_V\right )+
\sigma_{AA}\left ( 1+{4\alpha_s(m_t)\over{3\pi}}K_A\right ).
\end{eqnarray}
 For the  exact expressions of  $K_{V,A}$ we refer to Ref.\cite{Zttcor}.
$\sigma_{VV}$ and $\sigma_{AA}$ are the contributions to the total cross
section coming respectively from the pure vectorial part and the pure axial
part of the top couplings to $\gamma$, $Z$. 
We have not applied any 
correction to the interference term $\sigma_{VA}$ which vanishes when
integrated over the full phase space. Our treatment of QCD corrections is in
any case exact only in the narrow width approximation 
 for the total cross section with no cuts.
In all other cases it must be considered as a rough estimate of the
most important QCD corrections. In many
cases this has however proved to be a reasonable approximation,
probably just because the error
on the corrections  reflects upon a much smaller error on the cross sections.

For the numerical part we have used the $\rm G_{\mu}$-scheme 
\begin{eqnarray}
s_{_W}^2 = 1 - {{\wm^2}\over {\zm^2}}, \qquad 
g^2 &=& 4{\sqrt 2}\gf\wm^2
\end{eqnarray}
and the  input masses  $m_Z=91.1888$ GeV, $m_W=80.23$ GeV. We have chosen
 $m_t=180$ GeV and for $m_b$ the running mass value $m_b=2.7$ GeV. The top mass 
 we have  used  is actually  1~$\sigma$
apart from the actual measured central value\cite{top}, but all our 
conclusions are of   course insensitive to differences of the order of
the experimental error in the top mass. 
For the strong corrections to  Z, W and top decay widths
and vertices  we have used
$\alpha_s(m_Z)=$ 0.123 and evoluted it to the appropriate scales.

We have moreover implemented the following general set of cuts :
\bd
\item{-} jet(quark) energy $> 3$ GeV;
\item{-} lepton energy $> 1$ GeV;
\item{-} jet-jet invariant mass $> 10$ GeV;
\item{-} lepton-beam angle $> 10^{\circ}$;
\item{-} jet-beam angle $> 5^{\circ}$; 
\item{-} lepton-jet angle $> 5^{\circ}$.
\ed

Other cuts specific to particular studies will be described in the following.

Some results concerning top cross sections have been compared with computations
performed by the Grace group\cite{kuri} and we have found a substantial 
agreement.

\section {Top in the continuum}

The future $e^+e^-$ colliders will produce a great number of top top
events so that they  can also be regarded as  top factories: 
at 500 GeV the cross
section $\rm \sigma(t\bar t)$ is of the order of .5 pb, which corresponds
to about $10^4$ events per year for an integrated luminosity of 20 $\rm
fb^{-1}$.

We consider now two specific final states: $e^+e^-\ra \mu\bar\nu_{\mu} u\bar d
b \bar b$ and $e^+e^-\ra e \bar\nu_e u\bar d b \bar b$. We have computed at 
500 GeV and at 800 GeV the cross sections
 due to the signal 
diagrams only, and the irreducible backgrounds due to 
all (207 for the $\mu$ 
and 416 for the $e$) other diagrams and their interference with the signal.
The signal diagrams correspond to the first diagram of the second line of 
fig.~1 when in it only the first diagram of eq.~(7) is used.   
The results  are given in table~\ref{tabcstop}

\begin{table}[hbt]\centering
\begin{tabular}{|c|c|c|c|} 
\hline
\rule [-0.25 cm]{0 cm}{0.75 cm}
$\sqrt{s}$ GeV& channel & $t\bar t$ signal (fb) & background (fb) \\ \hline
\hline
\rule [-.25 cm] {0 cm} {0.75 cm} 
500 & $\mu \bar \nu u\bar d b\bar b$ & 19.850(4) & 0.736(3) \\ \cline{2-2}
\cline{4-4}
\rule [-.25 cm] {0 cm} {0.75 cm} 
  & $e \bar \nu u\bar d b\bar b$  &  & 0.778(5) \\ \hline
\rule [-0.25 cm]{0 cm}{0.75 cm}
800 &$\mu \bar \nu u\bar d b\bar b$ & 10.700(2) & 1.007(4) \\ \cline{2-2}
\cline{4-4}
\rule [-0.25 cm]{0 cm}{0.75 cm}
 &$e \bar \nu u\bar d b\bar b$ &  & 1.21(2) \\ \hline
\end{tabular}
\caption{Cross section for the processes $e^+ e^-~\ra\;\mu\bar
\nu u\bar d b\bar b$ and $e^+ e^- \ra\; e\bar \nu u\bar d b\bar
b$.}
\label{tabcstop}
\end{table}
\vspace{1 truecm}
Here and in the following  we report between parenthesis 
the  statistical integration errors on the last digit of the result.
For instance 19.850(4) has to be intended as $19.850\pm  0.004$.

These values have been obtained taking into account b masses, the full set
of diagrams, ISR, BST, NQCD corrections for the decay vertices of
the W's, the Z, tops and also for Z($\gamma$)tt vertex, as already explained.

Let us first discuss the relevance of these corrections on the specific example
of $e^+e^-\ra \mu\bar\nu_{\mu} u\bar d b \bar b$. In table~\ref{tabcstt500}
we report all results of this study. 
\begin{table}[hbt]\centering
\begin{tabular}{|c|c|c|}
\hline
\rule [-0.25 cm]{0 cm}{0.75 cm}
$e^+ e^- \ra \mu\bar \nu u\bar d b\bar b$ & $t\bar t$ signal (fb)& background
(fb)\\ \hline \hline
\rule [-0.25 cm]{0 cm}{0.75 cm}
$\rm NWA$                        & 18.880(3) & 0.848(3)\\ \hline
\rule [-0.25 cm]{0 cm}{0.75 cm}
$\rm Born$                       & 18.286(3) & 0.824(3)\\ \hline
\rule [-0.25 cm]{0 cm}{0.75 cm}
$\rm ISR$                        & 17.419(3) & 0.750(3)\\ \hline
\rule [-0.25 cm]{0 cm}{0.75 cm}
$\rm ISR\; NQCD^*$                   & 17.188(3) & 0.753(3)\\ \hline
\rule [-0.25 cm]{0 cm}{0.75 cm}
$\rm ISR\; NQCD^*\; BST$               & 17.303(3) & 0.731(3)\\ \hline
\rule [-0.25 cm]{0 cm}{0.75 cm}
$\rm ISR\; NQCD^*\; SBAND$             & 17.308(3) & 0.728(3)\\ \hline
\rule [-0.25 cm]{0 cm}{0.75 cm}
$\rm ISR\; NQCD^*\; BST\; m_b$         & 17.352(3) & 0.736(3)\\ \hline
\rule [-0.25 cm]{0 cm}{0.75 cm}
$\rm ISR\; NQCD\; BST\; m_b$& 19.850(4) &  0.736(3)\\ \hline
\end{tabular}
\caption[] { Cross section for the process $e^+ e^- \ra\; \mu\bar
\nu u\bar d b\bar b$ at $\sqrt{s}=500$ GeV for different sets of
approximations}
\label{tabcstt500}
\end{table}

The first result (NWA) reproduces what one would obtain using on shell 
calculation of $t\bar t$ production with subsequent decays of the tops to $bW$ 
and on shell decay of the $W$'s. Instead of computing it this way, we have used
the fact that \pd results can be obtained as a limit
of the corresponding off shell diagrams. One just performs the following
substitution for unstable particles propagator denominators:
\be
p^2-m^2+im\Gamma \qquad \ar \qquad (p^2-m^2+im\gamma)\:
\sqrt{\frac{\Gamma}{\gamma}}
\ee
 and then let  $\gamma \ar 0$. We have performed such limit numerically: 
we have used  ${\gamma}=10^{-3}\Gamma$ and we have checked that the result
does not vary diminishing $\gamma $. 

The Born approximation signal result
shows that taking into account the "off-shellness" of the unstable particles
produces a non-negligible variation of some percent. 

The introduction of ISR  decreases the $t\bar t$ signal with respect to 
Born only. This is at first surprising as around 500 GeV $t\bar t$ 
Born cross section is decreasing with energy and ISR reduces the effective
energy. In effect we will see in the following that
beamstrahlung raises the cross section. We have also checked that at 800 GeV
ISR  also raises it. The point is that, as we have verified, at 500 GeV
there is a probability of about 5\% that ISR reduces the energy of the hard
scattering below the $t\bar t$ threshold. This justifies the decrease, but
it also points out that there is a relatively high probability that ISR
takes the energy back around the threshold. For precision calculations the
corrections at threshold should therefore be included also at 500 GeV.

In the subsequent results of table~\ref{tabcstt500} we indicate with ${\rm
NQCD^*}$ the corrections to $t$'s, $W$'s and $Z$ decay vertices, with NQCD 
 these same corrections together with that of $Z(\gamma)t\bar t$ vertex.
Introducing ${\rm NQCD^*}$ lowers the signal. This fact can be easily
understood if we compute the same variation in \pd approximation. 
The signal has two
W's decaying to $\mu\bar\nu$ and $u \bar d$ respectively. The result
can be approximated by $t\bar t$ cross section multiplied by the two branching
$\Gamma_{W\ar \mu\bar\nu}/\Gamma_{W tot}$ and $\Gamma_{W\ar
u\bar d}/\Gamma_{W tot}$. Correcting with ${\rm NQCD^*}$ corresponds
to multiplying each hadronic $\Gamma_{q\bar q'}$ by $ 1+\alpha/\pi$.
$\Gamma_{W tot}$ has therefore to be multiplied by  $ 1+2\alpha/3\pi$.
The two branchings (and hence approximately also the signal cross section)
get in conclusion a factor $ (1+\alpha/\pi)/(1+2 \alpha/ 3 \pi)^2$ which
explains the difference between ISR and ISR ${\rm NQCD^*}$ entries.

Taking into account beamstrahlung (BST) 
results in a raise of the order of  half per cent  for the signal and
of almost three per cent for the background. Of course such a variation can be 
energy and process dependent, exactly as it happens with ISR. 
The program CIRCE that we have used allows for different parametrizations
of beamstrahlung effects. The one we have normally used  is TESLA. We have
tried also SBAND and from the result in  table~\ref{tabcstt500}
 one sees  that it does 
not introduce sensible differences from TESLA.

The introduction of the (running) b mass produces only a few permill variation.

It has to be noticed  an important
contribution (more than 10\%) to $\rm \sigma(signal)$ coming from the NQCD 
correction of the vertex $Z(\gamma)t\bar t$. It may of course be questionable
the way we have introduced such a correction, but in view of its numerical
relevance, we have considered that even a very crude approximation was
better than neglecting it, expecially in considerations of the relative
importance of signal and background.

Having briefly discussed the various approximations, let us now come  to
the main point  of considering  the irreducible background
due to all diagrams contributing to a final state. From table~\ref{tabcstop} 
 one can conclude that the background is important as it represents a 
correction of about 4\% to the signal.  Nevertheless such a statement has to be
 analyzed in more detail with the help of figs.~2 - 4. 
In fig.~2 and 3 it is reported
the invariant mass distribution of the top candidate in the channel 
$ e^+e^-\ra \mu\bar\nu u \bar d b \bar b$. We have not reported the
similar figure for $ e^+e^-\ra e\bar\nu u \bar d b \bar b$ as the conclusions
are practically identical. From table~\ref{tabcstop} one can easily see in fact
that the difference between the two channels ($\mu$ and e) is sizeable 
with respect to the backgrounds themselves, but it amounts to only a few 
permill of the signal. 

Given such a final state, one is faced with the problem of trying to identify
the particles which might come from a  top decay. 
In any single event one can try to measure
the invariant mass of three particles which could form the top.  
 Three of them are the muon, its neutrino and the $\bar b$. 
From this triplet it is experimentally  difficult to reconstruct the
invariant mass, because the neutrino momentum
can be deduced only from missing momentum, to which also ISR and
BST contribute. 
The best strategy is to try to identify 
the three quarks forming the top. We assume that there is b tagging,
and we require that both b's are identified,
so that they are separated from the other two quarks.
One cannot  distinguish between a $b$ and a $\bar b$. The only
possibility is to form two invariant masses with one of the two b's and
the other  light quarks. Considering the distribution on the sum of these two
invariant masses, it will be possible to measure the mass of the top.
Once this has been measured, one can look for the nearest 
to the expected top mass between the two invariant masses.
We refer to  this one as the mass of the top candidate, and we have plotted
its distribution. In fig.~2 one can see  the differences among
 this distribution, the one due to the signal diagrams and that due to
 background diagrams. Also this last distribution peaks at the top mass.
 This is of course due to the many diagrams which are "single resonant",
in which one of the two top propagators  can go on mass shell.
 We have also computed the distribution after some cuts have been applied to
 partially eliminate the background (dotted lines).
 The cuts we have imposed are:
\be
\rm   |m(ud)-m_W|<20\:GeV \qquad |m(b\bar b)-m_Z|>20\:GeV \label{cut1}
\ee
One can see that these cuts reduce in fact the background by about a factor two,
 but they do not affect the peak at the top mass. 

In fig.~3 we have reported on a linear scale and on the neighborhood 
of the peak the three curves relative
to full process, signal and full process after cuts, both at 500 GeV and 
at 800 GeV. 
From these curves one concludes that there doesn't seem to be 
any difference in the location of the maximum, but there are some appreciable
differences in the height between signal and total distributions.
The cuts (\ref{cut1}) do not seem to help in this region.

We have also studied the angular dependence of the top candidate. This is
reported in fig.~4. If one compares the dashed and full curves, one
notice that both at 500 GeV and at 800 GeV, there is a difference in the
angular distribution between signal and total calculations. This is
particularly relevant in the forward $e^+$ direction. In order to try to reduce
the contribution of the irreducible background, we have studied the total
distribution with the cuts~(\ref{cut1}) and an additional cut
 on the mass of the top candidate $m_{tc}$: 
\be
|m_{tc}-m_t|<40\:GeV \label{cut2}
\ee  
The net result of such cuts, as it can be seen from the dotted curves, is
 to effectively reduce the  irreducible background in the
 forward direction, but a mild distortion of the total curve, with
respect to the signal one, remains also after the cuts.

The study of possible cuts is of course not exhausted by this short discussion,
and the distributions presented are just an example of  how one can handle 
this problem with  complete calculations.

\section {WWZ and its background}
In this section we study the semileptonic processes in which there are
no $b$'s in the final state. 
As we have already mentioned in section 2, one of the reasons why  these 
processes are of interest is that  they allow to study WWZ formation and 
decay where it will be possible to test the quartic gauge coupling.

The diagrams one has to deal with are essentially
those for the full calculations of the preceding section. The main difference
comes from the fact that those diagrams 
which were double or single resonant for top propagators, now do not
have the top or antitop as an intermediate state, but a lighter quark.
Their contribution is therefore greatly reduced and the corresponding total 
cross section decreases by an amount comparable to the size
 of  $t\bar t$ signal diagrams.

The most important contribution to the cross section for the processes at hand
comes from the 15 diagrams which correspond to WWZ production and decay.
These have three resonant vector boson propagators and we will call them
signal diagrams. The remaining ones can be divided in
 double, single and non resonant parts.
 
The results we will present will take into account ISR, BST, NQCD.
We want to point out  that for these processes the difference between
calculations taking or not into account beamstrahlung is important.
In fact, if we take as an example the process $e^+e^-\ra \mu \bar \nu_\mu 
u \bar d s \bar s$, the  full cross section  is  .18480(9) fb without
BST and .18033(9) fb with BST. That for WWZ signal only is  .17739(3) fb
without BST and .17303(3) fb with BST. So the difference is of the order of
two percent. If we compare this with the analogous differences in the case
of top signal (half percent) and top background (almost three percent), we
realize once again that BST effects are process (and cuts) dependent.
Similarly to what happens  for ISR, one cannot just ignore  or approximate them
with an overall factor. 

In table~\ref{cswwz500}  we present the cross sections for the full processes, 
those
computed taking into account signal diagrams  only, and those
computed via the \pd approximation. This
last result can also be obtained  taking the narrow width
approximation limit of the signal diagrams. We have taken such a limit
numerically as we have done for the top case. 

\begin{table}[hbt]\centering
\begin{tabular}{|c|c|c|c|}
\hline
\rule [-0.25 cm]{0 cm}{0.75 cm}
process                         & WWZ NWA (fb) & WWZ signal (fb) & complete (fb)
\\
\hline \hline
\rule [-0.25 cm]{0 cm}{0.75 cm}
$\mu \bar \nu u\bar d c\bar c$  & 0.13836(2)   & 0.13464(2)      & 0.16218(9) \\
\cline{1-1} \cline{4-4}
\rule [-0.25 cm]{0 cm}{0.75 cm}
$e \bar \nu u\bar d c\bar c$    &            &             & 0.1783(2) \\
\hline
\rule [-0.25 cm]{0 cm}{0.75 cm}
$\mu \bar \nu u\bar d s\bar s$  & 0.17780(3) & 0.17303(3)  & 0.1803(1) \\
\cline{1-1} \cline{4-4}
\rule [-0.25 cm]{0 cm}{0.75 cm}
$e \bar \nu u\bar d s\bar s$    &            &             & 0.2117(2) \\
\hline
\rule [-0.25 cm]{0 cm}{0.75 cm}
$\mu \bar \nu u\bar d u\bar u$  & 0.12815(2) & 0.12469(2)  & 0.1512(1) \\
\cline{1-1} \cline{4-4}
\rule [-0.25 cm]{0 cm}{0.75 cm}
$e \bar \nu u\bar d u\bar u$    &            &             & 0.1758(3) \\
\hline
\rule [-0.25 cm]{0 cm}{0.75 cm}
$\mu \bar \nu u\bar d d\bar d$  & 0.16468(3) & 0.16025(3)  & 0.16733(9) \\
\cline{1-1} \cline{4-4}
\rule [-0.25 cm]{0 cm}{0.75 cm}
$e \bar \nu u\bar d d\bar d$    &            &             & 0.1941(1)  \\ 
\hline
\end{tabular}
\caption[]{Cross section for the processes $e^+e^-\ra l\bar \nu_l + 4$ light
quarks ($l=\mu,e$) at $\sqrt{s}=500$ GeV}
\label{cswwz500}
\end{table}

The difference between the full calculation and on shell (NWA) approximation is
indeed remarkable. Even finite width effects are of some importance as they
determine a variation of some percent between signal and NWA. 
We notice that the background is much higher in a process with an up-type quark
pair than in the analogous one with down-type. For instance 
$\mu \bar \nu u\bar d c\bar c$  background is .0275fb while the 
$\mu \bar \nu u\bar d s\bar s$ is .0073fb. As the dominant contributions to
background are the double resonant ones, the previous consideration shows that
for the set of cuts we are using, most background comes from diagrams with two
resonant W's and a $\gamma$ converting in a quark-antiquark pair. Diagrams with
 one resonant W and the pair coming from Z decay would in fact produce 
an opposite behaviour and
the numerical difference between the two backgrounds approximately corresponds
to that due to the different quark charges in $WW\gamma^*$ diagrams.

In presence of
such big irreducible background, it is necessary to introduce other cuts 
in order to
try to isolate WWZ production.
We have performed detailed studies of possible cuts and we present them in 
figs.~(5-7). Obviously  they are at parton
level, but just this fact helps in understanding the theoretical differences
and the size of the errors associated with the various approximations.
In practice, before considering hadronization, detector simulations and so
on, it is important to study whether  the full
calculation is needed for a certain set of cuts or, for instance,  signal 
diagrams represent a viable approximation. This may indeed be understood 
performing a comparison at parton level.

We have implemented  different cuts on invariant masses of the pairs of 
particles which should come from vector bosons, and computed the cross sections
for the various cuts with the full set of diagrams, with only the fifteen
signal diagrams,
and with \pd or narrow width approximation.
It is obvious that the cross sections computed with the last approximation  
are  not sensible
to such kind of cuts, as the vector bosons are always considered on shell.
 In figs.~(5-8) the dotted lines refer to this case.
For the above reason they are just straight lines and they
are reported just as a reference to show where the on shell cross section lies.
The other two kinds of lines represent signal (dashed) and complete (continuous)
cross sections as a function of  cuts.

In order to reduce the contribution of non WWZ diagrams 
(the so called irreducible background), one would like to constrain the
invariant masses formed by the appropriate final particles to be as near as 
possible to those of two W's and a Z. The momentum of the
neutrino is not directly measurable, and therefore the neutrino lepton 
invariant mass is difficult to measure. For this reason we first 
consider some cuts which act only on the quarks. 
We intend somehow to force two quarks to reproduce a W and the other two to 
reproduce a Z. As the four quarks are practically indistinguishable, we compute
the invariant mass of any pair of quarks. We then
accept an event if out of the three couples of pairs of quarks, one at least
has a pair whose invariant mass is around $M_W$ and the other pair around $M_Z$.
In  figs.~5-6 are reported the cross sections for such kind
of cuts as a function of $M_{cut}$, which represents the size of the window 
around $M_W$ and $M_Z$ in which events are accepted: the quarks are required 
to form at least one couple of  pairs whose 
invariant masses $m_i$ ($i=1,2$) satisfy the conditions $|M_V -m_i| < M_{cut}$ 
($V=W,Z$). Some comments are in order. Both figures show that the signal
contribution  to the cross section (dashed line) is lower that the NWA even for
very loose cuts or for the cross section without cuts (see table~\ref{cswwz500}). 
This implies
that one cannot rely on the narrow width approximation. The fact that the full
calculation at high $M_{cut}$ approaches the NWA for the $\mu \bar \nu u \bar d
s \bar s $ case (fig.~5) is probably just casual. It does not happen
for $e \bar \nu u \bar d s \bar s $. The curves of
fig.~6  show that for the electron case the difference between signal
and total process is extremely relevant. It grows with the energy and
the cuts we have imposed can greatly reduce the difference but not suppress it.
For the muon case an $M_{cut}$ of about 10 GeV is on the contrary sufficient
to make total and signal cross section practically coincide. The loss in
event number is however of the order of ten percent.

As we are interested primarily in WWZ physics, it is important to find out
whether further cuts may allow  also in the electron case to further suppress
the background. We have therefore studied the possibility of imposing a cut
also on the invariant mass formed by the electron and the neutrino.
In fig.~7 the cross sections for two cases are reported:
\bd
\item{a)}
quarks are required to form two pairs whose invariant masses $m_i$ ($i=1,2$)
satisfy the conditions $|M_V -m_i| < 15$ GeV, $V=W,Z$
and $e^-\bar \nu_e$ are required to form an invariant mass m such that $|M_W -
m| <M_{cut}$.
\item{b)}
quarks are required to form two pairs whose invariant masses $m_i$ ($i=1,2$)
satisfy the conditions $|M_V -m_i| < M_{cut}$, $V=W,Z$
and $e^-\bar \nu_e$ are required to form an invariant mass m such that $|M_W -
m| <M_{cut}$.
\ed
In the latter case the simultaneous variation of $M_{cut}$ is studied, while
in the first we explore the possibility of having different cuts on the 
leptonic and quark system.
As explained before the determination of $\nu\: e$ invariant mass is affected
by great experimental errors. For such a reason 
in fig.~7 we have reported
the cross sections for this invariant mass and for the more realistic case
in which the cut is imposed on the invariant mass formed with electron and
reconstructed neutrino four momenta. We assign in the latter  case
  all missing
three-momentum $\bar p_{mis}$ to the neutrino and take its energy 
to be equal to $|\bar p_{mis}|$.

The conclusions that can be drawn from fig.~7 are almost self evident:
Comparing fig.~6 with fig.~7, one can verify that even
introducing a very loose cut as $M_{cut}$=60~GeV on $e^-\bar\nu_e$ invariant
mass reduces significantly the difference between signal and total cross 
sections.
This  means that the contribution of exchanged diagrams which were
responsible of the great difference between  the electron and the muon case
is sensitive to this kind of cut. On the other hand, fig.~7 shows that
if the cuts have to be so stringent to reduce the difference between signal and
total cross section to the order of the percent, one looses  more than one
third
of the event number, as compared to the NWA. This conclusion is important
in view of the fact that such WWZ processes have a low cross section:
at 500 GeV $\rm \sigma(WWZ)$ is of the order of 40fb, which corresponds
to a total of 800 events per year for an integrated luminosity of 20 $\rm
fb^{-1}$.

The reactions we have just examined in figs.~5-7 have been used
as a case study for the cuts, but in reality they cannot be directly measured.
This is due to the fact that quark flavors cannot be disentangled
experimentally.
We examine therefore in fig.~8 the more interesting physical case
in which we sum over all reactions involving $\mu$'s. To this end we have
computed for all different cuts the cross sections of the four reactions:
\[
  e^+e^-\ra \mu^- \bar \nu_\mu u \bar d s \bar s \qquad
  e^+e^-\ra \mu^- \bar \nu_\mu u \bar d c \bar c  \qquad
  e^+e^-\ra \mu^- \bar \nu_\mu u \bar d u \bar u \qquad
  e^+e^-\ra \mu^- \bar \nu_\mu u \bar d d \bar d  
\]

In the plot it is reported, both for WWZ diagrams  and for the complete
calculation, 
the sum of these cross sections multiplied by a factor
4. This accounts for the reactions in which one has $\mu^+  \nu_\mu \bar u d$ 
instead of $\mu^- \bar \nu_\mu u \bar d$ and for those in which one 
has $c \bar s$ (or $\bar c s$) instead of $u \bar d$ (or $\bar u  d$).
Under this exchanges the sum of the four reactions above remains 
the same for the present set of cuts. The dashed and continuous lines of 
fig.~8 give therefore
the exact cross section (at order $\alpha^6$) as a function of $M_{cut}$ for 
all processes with one  muon, four quarks  and no b's in the final state.

We are implicitly  assuming in these considerations that, in order to 
exclude the overwhelming background from $t \bar t$ production and decay,
b-tagging will be used to exclude all events with at least one tagged b. 
This is the reason why we do not sum on WWZ events
with b's in the final state. With actual
b-tagging techniques, there is a high probability $P_{c\ra b}$
(say of the order of thirty per cent) that a $c$ be misidentified as a $b$.
To take this into consideration,  one 
should  multiply by the appropriate reduction factors $1-P_{c\ra b}$,
$(1-P_{c\ra b})^2$, $(1-P_{c\ra b})^3$
the contributions with one, two or three $c$'s
to the the sums reported in fig.~8.
This would lead to a decrease of the curves of about 25\% for $P_{c\ra b}=.3$.

True  events  with two b's in the final state, 
which are dominated by top top signal  can be misidentified and 
become an important background to WWZ physics 
as there is a finite probability $P_{b\ra \slaar b}=1-P_{b\ra b}$ that a b
may not be recognized as such. 
We have computed such a background as a function of $M_{cut}$. The chain dash
and chain dot curves of fig.~8 correspond  
to the contribution of top top signal and background (to $t\bar t$ events) 
respectively. We have here assumed $P_{b\ra \slaar b}=.2$ and summed over 
the processes 
\[
  e^+e^-\ra \mu^- \bar \nu_\mu u \bar d b \bar b \qquad
  e^+e^-\ra \mu^- \bar \nu_\mu c \bar s b \bar b  \qquad
  e^+e^-\ra \mu^+      \nu_\mu \bar u d b \bar b \qquad
  e^+e^-\ra \mu^+      \nu_\mu \bar c s b \bar b.
\]

The first consideration on this background coming from top top events is
that even an imperfect b tagging is of considerable help in strongly 
reducing the great number of these events. It has to be remarked that
this background depends strongly on the applied
$M_{cut}$. If one adopts the strategy of applying a severe $M_{cut}$ 
of the order of 10 GeV, it is reduced to about 1/6 of WWZ signal.
If on the other hand a milder $M_{cut}$ is used or if $P_{b\ra\slaar b}$ is
greater than what we used, it may become comparable to the signal itself. 
In such a case, other cuts may still be used. For instance we have verified
that  by requiring all four possible three jets
invariant masses to be
out of a window of 40 GeV around the top mass,
one reduces the contribution of top top events to be less than that of the 
other diagrams for $b\bar b$ final states (chaindot curve in fig.~8),
without sensible loss of WWZ signal.

\section{Conclusions}
We have computed tree level semileptonic six fermion processes which will 
become relevant for the future $e^+e^-$ linear colliders. Our results
concern at present the infinite Higgs mass limit.
The helicity amplitude method used~\cite{phact}\cite{method} has allowed us
to build up a program (SIXPHACT) which is able to produce technically precise 
results in  reasonable computer time. This was far from obvious at the start,
given the complexity of treating a 6 body final state and the great number of
complex Feynman diagrams. Speed and precision are essential to produce
very reliable differential cross sections. These allow to exploit the 
true potentiality of  six fermion complete calculations: it is just studying
distributions that one can understand the differences from the usual 
\pd approximation and which cuts have to be imposed to enhance signals with
respect to irreducible background.

The main drawback when dealing with complete tree level matrix elements
is how to account for radiative corrections. 
In this respect on shell computations folded with decays are in a much better
shape. On the other hand they completely lack the contributions to the
physical final states of a large class of diagrams and  they miss finite
width and spin correlation effects. In our opinion the two ways of approaching
 the physical problem are somehow complementary and must both be pursued. 
We have included initial state radiation and beamstrahlung.
QCD corrections are taken into account in a simplified (naive) way, exact
only in NWA for total cross sections without cuts.

We have given some examples  of phenomenological studies relevant
to top and WWZ physics at 500 and 800 GeV.
We have in particular found that single resonant background contributions are
difficult to get rid of when studying invariant mass or angular 
top distributions. For WWZ it seems that \pd approximation is not viable,
and that to get rid of the irreducible background with the cuts
we have tried, one looses about 10\%
of events with a muon in the final state and more than 30\% for an electron 
at 500 GeV.
The enormous amount of background coming from $t \bar t$ production seems
to be completely disposed of with appropriate b-tagging and cuts. The effect
of cuts does not influence the signal, while spurious b-tagging of signal
events  may result in a decrease of the order of 25\%.

\section*{Acknowledgments}
 We wish to thank  Yoshimasa Kurihara and Roberto Pittau for useful
 discussions. Comparisons with Y. Kurihara and F. Yuasa are also 
 gratefully aknowledged.

\vfill\eject

\vfill\eject
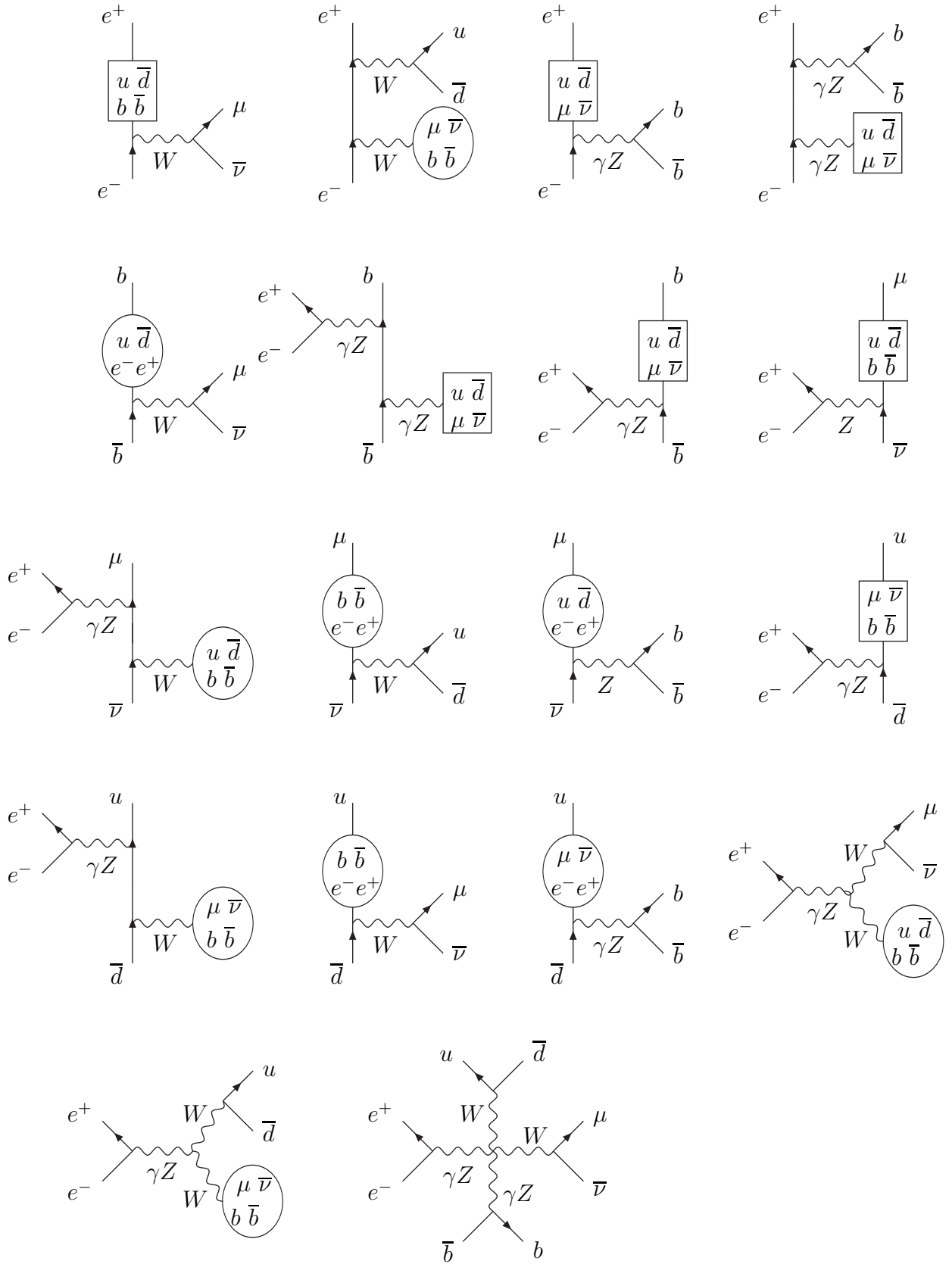
\begin{figure}%[p]

%le misure sono in points. 1 point=.35mm. 
%quindi approssimativamente 3=1mm, 30=1cm. Cosi' il size della picture
%(che mi sembra circa il massimo possibile e' circa 14x21
\begin{picture}(420,630)(0,0)

%diag 1
\SetOffset(50,620)
%12
\Boxc(0,0)(24,30)
\Text(-8,12)[lt]{$u$ $\ol d$}
\Text(-8,-12)[lb]{$b$ $\ol b$}
\Line (0,15)(0,34)
%\Line (0,-15)(0,-34)
\Text(-3,34)[rb]{$e^+$}
%\Text(-3,-34)[rt]{$1$}
\SetOffset(50,596)
%linea verticale con eventuali impulsi
\ArrowLine(0,-20)(0,9)
%\Text(-5,20)[rb]{$1$}
\Text(-5,-20)[rt]{$e^-$}
% gamma,Zeta ff secondario
\Photon(0,0)(30,0){2}{3}
\Text(10,-15)[lb]{$W$}
\ArrowLine(30,0)(45,15)
\Text(50,15)[l]{$\mu$}
\Line (30,0)(45,-15)
\Text(50,-15)[l]{$\ol \nu$}

%diag 2
\SetOffset(160,634)
%linea verticale con eventuali impulsi
\ArrowLine(0,-20)(0,20)
\Text(-5,20)[rb]{$e^+$}
%\Text(-5,-20)[rt]{$1$}
% gamma,Zeta ff secondario
\Photon(0,0)(30,0){2}{3}
\Text(10,-15)[lb]{$W$}
\ArrowLine(30,0)(45,15)
\Text(50,15)[l]{$u$}
\Line (30,0)(45,-15)
\Text(50,-15)[l]{$\ol d$}
\SetOffset(160,594)
%17
%linea verticale con eventuali impulsi
\ArrowLine(0,-20)(0,20)
%\Text(-5,20)[rb]{$P$}
\Text(-5,-20)[rt]{$e^-$}
% gamma ff primario piccolo
\Photon(0,0)(30,0){2}{3}
\Text(10,-14)[lb]{$W$}
\Oval(45,0)(18,15)(0)
\Text(37,12)[lt]{$\mu$ $\ol \nu$}
\Text(37,-12)[lb]{$b$ $\ol b$}

%diag 3
\SetOffset(270,620)
%12
\Boxc(0,0)(24,30)
\Text(-8,12)[lt]{$u$ $\ol d$}
\Text(-8,-12)[lb]{$\mu$ $\ol \nu$}
\Line (0,15)(0,34)
%\Line (0,-15)(0,-34)
\Text(-3,34)[rb]{$e^+$}
%\Text(-3,-34)[rt]{$1$}
\SetOffset(270,596)
%linea verticale con eventuali impulsi
\ArrowLine(0,-20)(0,9)
%\Text(-5,20)[rb]{$1$}
\Text(-5,-20)[rt]{$e^-$}
% gamma,Zeta ff secondario
\Photon(0,0)(30,0){2}{3}
\Text(10,-15)[lb]{$\gamma Z$}
\ArrowLine(30,0)(45,15)
\Text(50,15)[l]{$b$}
\Line (30,0)(45,-15)
\Text(50,-15)[l]{$\ol b$}

%diag 4
\SetOffset(380,634)
%linea verticale con eventuali impulsi
\ArrowLine(0,-20)(0,20)
\Text(-5,20)[rb]{$e^+$}
%\Text(-5,-20)[rt]{$1$}
% gamma,Zeta ff secondario
\Photon(0,0)(30,0){2}{3}
\Text(10,-15)[lb]{$\gamma Z$}
\ArrowLine(30,0)(45,15)
\Text(50,15)[l]{$b$}
\Line (30,0)(45,-15)
\Text(50,-15)[l]{$\ol b$}
\SetOffset(380,594)
%16
%linea verticale con eventuali impulsi
\ArrowLine(0,-20)(0,20)
%\Text(-5,20)[rb]{$P$}
\Text(-5,-20)[rt]{$e^-$}
% gamma ff primario piccolo
\Photon(0,0)(30,0){2}{3}
\Text(8,-14)[lb]{$\gamma Z$}
\Boxc(42,0)(24,30)
\Text(34,12)[lt]{$u$ $\ol d$}
\Text(34,-12)[lb]{$\mu$ $\ol \nu$}

%%%%%%%%%%%%%%%%%%%%%%%%%%%%

%diag 5
\SetOffset(50,490)
%13
\Oval(0,0)(18,15)(0)
\Text(-8,12)[lt]{$u$ $\ol d$}
\Text(-11,-12)[lb]{$e^- e^+$}
\Line (0,18)(0,34)
%\Line (0,-18)(0,-34)
\Text(-3,34)[rb]{$b$}
%\Text(-3,-34)[rt]{$1$}
\SetOffset(50,464)
%linea verticale con eventuali impulsi
\ArrowLine(0,-20)(0,8)
%\Text(-5,20)[rb]{$1$}
\Text(-5,-20)[rt]{$\ol b$}
% gamma,Zeta ff secondario
\Photon(0,0)(30,0){2}{3}
\Text(10,-15)[lb]{$W$}
\ArrowLine(30,0)(45,15)
\Text(50,15)[l]{$\mu$}
\Line (30,0)(45,-15)
\Text(50,-15)[l]{$\ol \nu$}

%diag 6
\SetOffset(175,504)
%linea verticale con eventuali impulsi
\ArrowLine(0,-20)(0,20)
\Text(-5,20)[rb]{$b$}
%\Text(-5,-20)[rt]{$1$}
% gamma,Zeta ff secondario
\Photon(0,0)(-30,0){2}{3}
\Text(-23,-15)[lb]{$\gamma Z$}
\ArrowLine(-30,0)(-45,15)
\Text(-50,15)[r]{$e^+$}
\Line (-30,0)(-45,-15)
\Text(-50,-15)[r]{$e^-$}

\SetOffset(175,464)
%linea verticale con eventuali impulsi
\ArrowLine(0,-20)(0,20)
%\Text(-5,20)[rb]{$P$}
\Text(-5,-20)[rt]{$\ol b$}
% gamma ff primario piccolo
\Photon(0,0)(30,0){2}{3}
\Text(8,-14)[lb]{$\gamma Z$}
\Boxc(42,0)(24,30)
\Text(34,12)[lt]{$u$ $\ol d$}
\Text(34,-12)[lb]{$\mu$ $\ol \nu$}

%diag 7
\SetOffset(315,490)
%12
\Boxc(0,0)(24,30)
\Text(-8,12)[lt]{$u$ $\ol d$}
\Text(-8,-12)[lb]{$\mu$ $\ol\nu$}
\Line (0,15)(0,34)
%\Line (0,-15)(0,-34)
\Text(5,34)[lb]{$b$}
%\Text(-3,-34)[rt]{$1$}

\SetOffset(315,464)
%linea verticale con eventuali impulsi
\ArrowLine(0,-20)(0,11)
%\Text(-5,20)[rb]{$1$}
\Text(5,-20)[lt]{$\ol b$}
% gamma,Zeta ff secondario
\Photon(0,0)(-30,0){2}{3}
\Text(-23,-15)[lb]{$\gamma Z$}
\ArrowLine(-30,0)(-45,15)
\Text(-50,15)[r]{$e^+$}
\Line (-30,0)(-45,-15)
\Text(-50,-15)[r]{$e^-$}

%diag 8
\SetOffset(425,490)
%12
\Boxc(0,0)(24,30)
\Text(-8,12)[lt]{$u$ $\ol d$}
\Text(-8,-12)[lb]{$b$ $\ol b$}
\Line (0,15)(0,34)
%\Line (0,-15)(0,-34)
\Text(5,34)[lb]{$\mu$}
%\Text(-3,-34)[rt]{$1$}

\SetOffset(425,464)
%linea verticale con eventuali impulsi
\ArrowLine(0,-20)(0,11)
%\Text(-5,20)[rb]{$1$}
\Text(5,-20)[lt]{$\ol \nu$}
% gamma,Zeta ff secondario
\Photon(0,0)(-30,0){2}{3}
\Text(-23,-15)[lb]{$Z$}
\ArrowLine(-30,0)(-45,15)
\Text(-50,15)[r]{$e^+$}
\Line (-30,0)(-45,-15)
\Text(-50,-15)[r]{$e^-$}

%%%%%%%%%%%%%%%%%%%%%%%%%%%%

%diag 9
\SetOffset(50,364)
%linea verticale con eventuali impulsi
\ArrowLine(0,-20)(0,20)
\Text(-5,20)[rb]{$\mu$}
%\Text(-5,-20)[rt]{$1$}
% gamma,Zeta ff secondario
\Photon(0,0)(-30,0){2}{3}
\Text(-23,-15)[lb]{$\gamma Z$}
\ArrowLine(-30,0)(-45,15)
\Text(-50,15)[r]{$e^+$}
\Line (-30,0)(-45,-15)
\Text(-50,-15)[r]{$e^-$}
\SetOffset(50,334)
%linea verticale con eventuali impulsi
\ArrowLine(0,-20)(0,20)
%\Text(-5,20)[rb]{$P$}
\Text(-5,-20)[rt]{$\ol \nu$}
% gamma ff primario piccolo
\Photon(0,0)(30,0){2}{3}
\Text(10,-14)[lb]{$W$}
\Oval(45,0)(18,15)(0)
\Text(37,12)[lt]{$u$ $\ol d$}
\Text(37,-12)[lb]{$b$ $\ol b$}

%diag 10
\SetOffset(160,360)
%13
\Oval(0,0)(18,15)(0)
\Text(-8,12)[lt]{$b$ $\ol b$}
\Text(-11,-12)[lb]{$e^- e^+$}
\Line (0,18)(0,34)
%\Line (0,-18)(0,-34)
\Text(-3,34)[rb]{$\mu$}
%\Text(-3,-34)[rt]{$1$}
\SetOffset(160,334)
%linea verticale con eventuali impulsi
\ArrowLine(0,-20)(0,8)
%\Text(-5,20)[rb]{$1$}
\Text(-5,-20)[rt]{$\ol \nu$}
% gamma,Zeta ff secondario
\Photon(0,0)(30,0){2}{3}
\Text(10,-15)[lb]{$W$}
\ArrowLine(30,0)(45,15)
\Text(50,15)[l]{$u$}
\Line (30,0)(45,-15)
\Text(50,-15)[l]{$\ol d$}

%diag 11
\SetOffset(270,360)
%13
\Oval(0,0)(18,15)(0)
\Text(-8,12)[lt]{$u$ $\ol d$}
\Text(-11,-12)[lb]{$e^- e^+$}
\Line (0,18)(0,34)
%\Line (0,-18)(0,-34)
\Text(-3,34)[rb]{$\mu$}
%\Text(-3,-34)[rt]{$1$}
\SetOffset(270,334)
%linea verticale con eventuali impulsi
\ArrowLine(0,-20)(0,8)
%\Text(-5,20)[rb]{$1$}
\Text(-5,-20)[rt]{$\ol \nu$}
% gamma,Zeta ff secondario
\Photon(0,0)(30,0){2}{3}
\Text(12,-15)[lb]{$ Z$}
\ArrowLine(30,0)(45,15)
\Text(50,15)[l]{$b$}
\Line (30,0)(45,-15)
\Text(50,-15)[l]{$\ol b$}

%diag 12
\SetOffset(425,360)
%12
\Boxc(0,0)(24,30)
\Text(-8,12)[lt]{$\mu$ $\ol \nu$}
\Text(-8,-12)[lb]{$b$ $\ol b$}
\Line (0,15)(0,34)
%\Line (0,-15)(0,-34)
\Text(5,34)[lb]{$u$}
%\Text(-3,-34)[rt]{$1$}

\SetOffset(425,334)
%linea verticale con eventuali impulsi
\ArrowLine(0,-20)(0,11)
%\Text(-5,20)[rb]{$1$}
\Text(5,-20)[lt]{$\ol d$}
% gamma,Zeta ff secondario
\Photon(0,0)(-30,0){2}{3}
\Text(-23,-15)[lb]{$\gamma Z$}
\ArrowLine(-30,0)(-45,15)
\Text(-50,15)[r]{$e^+$}
\Line (-30,0)(-45,-15)
\Text(-50,-15)[r]{$e^-$}

%%%%%%%%%%%%%%%%%%%%%%%%%%%%

%diag 13
\SetOffset(50,244)
%linea verticale con eventuali impulsi
\ArrowLine(0,-20)(0,20)
\Text(-5,20)[rb]{$u$}
%\Text(-5,-20)[rt]{$1$}
% gamma,Zeta ff secondario
\Photon(0,0)(-30,0){2}{3}
\Text(-23,-15)[lb]{$\gamma Z$}
\ArrowLine(-30,0)(-45,15)
\Text(-50,15)[r]{$e^+$}
\Line (-30,0)(-45,-15)
\Text(-50,-15)[r]{$e^-$}
\SetOffset(50,204)
%linea verticale con eventuali impulsi
\ArrowLine(0,-20)(0,20)
%\Text(-5,20)[rb]{$P$}
\Text(-5,-20)[rt]{$\ol d$}
% gamma ff primario piccolo
\Photon(0,0)(30,0){2}{3}
\Text(10,-14)[lb]{$W$}
\Oval(45,0)(18,15)(0)
\Text(37,12)[lt]{$\mu$ $\ol\nu $}
\Text(37,-12)[lb]{$b$ $\ol b$}

%diag 14
\SetOffset(160,230)
%13
\Oval(0,0)(18,15)(0)
\Text(-8,12)[lt]{$b$ $\ol b$}
\Text(-11,-12)[lb]{$e^- e^+$}
\Line (0,18)(0,34)
%\Line (0,-18)(0,-34)
\Text(-3,34)[rb]{$u$}
%\Text(-3,-34)[rt]{$1$}
\SetOffset(160,204)
%linea verticale con eventuali impulsi
\ArrowLine(0,-20)(0,8)
%\Text(-5,20)[rb]{$1$}
\Text(-5,-20)[rt]{$\ol d$}
% gamma,Zeta ff secondario
\Photon(0,0)(30,0){2}{3}
\Text(10,-15)[lb]{$W$}
\ArrowLine(30,0)(45,15)
\Text(50,15)[l]{$\mu$}
\Line (30,0)(45,-15)
\Text(50,-15)[l]{$\ol \nu$}

%diag 15
\SetOffset(270,230)
%13
\Oval(0,0)(18,15)(0)
\Text(-8,12)[lt]{$\mu$ $\ol \nu$}
\Text(-11,-12)[lb]{$e^- e^+$}
\Line (0,18)(0,34)
%\Line (0,-18)(0,-34)
\Text(-3,34)[rb]{$u$}
%\Text(-3,-34)[rt]{$1$}
\SetOffset(270,204)
%linea verticale con eventuali impulsi
\ArrowLine(0,-20)(0,8)
%\Text(-5,20)[rb]{$1$}
\Text(-5,-20)[rt]{$\ol d$}
% gamma,Zeta ff secondario
\Photon(0,0)(30,0){2}{3}
\Text(10,-15)[lb]{$\gamma Z$}
\ArrowLine(30,0)(45,15)
\Text(50,15)[l]{$b$}
\Line (30,0)(45,-15)
\Text(50,-15)[l]{$\ol b$}

%diag 16
\SetOffset(380,220)
%e+e- entranti
\ArrowLine(0,0)(-15,15)
\Text(-20,15)[rb]{$e^+$}
\Line(0,0)(-15,-15)
\Text(-20,-15)[rt]{$e^-$}
\Photon(0,0)(30,0){2}{3}
\Text(7,-14)[lb]{$\gamma Z$}
\Photon(30,0)(45,25){2}{3}
\Text(25,15)[lb]{$W$}
\Photon(25,0)(45,-25){2}{3}
\Text(25,-20)[lt]{$W$}
\ArrowLine(45,25)(60,40)
\Text(65,40)[l]{$\mu$}
\Line (45,25)(60,10)
\Text(65,10)[l]{$\ol \nu$}
\SetOffset(440,195)
\Oval(0,0)(18,15)(0)
\Text(-8,12)[lt]{$u$ $\ol d$}
\Text(-11,-12)[lb]{$b$ $\ol b$}

%%%%%%%%%%%%%%%%%%%%%%%%%%%%

%diag 17
\SetOffset(50,90)
%e+e- entranti
\ArrowLine(0,0)(-15,15)
\Text(-20,15)[rb]{$e^+$}
\Line(0,0)(-15,-15)
\Text(-20,-15)[rt]{$e^-$}
\Photon(0,0)(30,0){2}{3}
\Text(7,-14)[lb]{$\gamma Z$}
\Photon(30,0)(45,25){2}{3}
\Text(25,15)[lb]{$W$}
\Photon(30,0)(45,-25){2}{3}
\Text(25,-20)[lt]{$W$}
\ArrowLine(45,25)(60,40)
\Text(65,40)[l]{$u$}
\Line (45,25)(60,10)
\Text(65,10)[l]{$\ol d$}
\SetOffset(110,65)
\Oval(0,0)(18,15)(0)
\Text(-8,12)[lt]{$\mu$ $\ol \nu$}
\Text(-11,-12)[lb]{$b$ $\ol b$}

%diag 18
\SetOffset(200,90)
%e+e- entranti
\ArrowLine(0,0)(-15,15)
\Text(-20,15)[rb]{$e^+$}
\Line(0,0)(-15,-15)
\Text(-20,-15)[rt]{$e^-$}
\Photon(0,0)(30,0){2}{3}
\Text(5,-16)[lb]{$\gamma Z$}
\Photon(30,0)(30,30){2}{3}
\Text(26,15)[rb]{$W$}
\ArrowLine(30,30)(15,45)
\Text(10,45)[rb]{$u$}
\Line(30,30)(45,45)
\Text(50,45)[lb]{$\ol d$}
\Photon(30,0)(60,0){2}{3}
\Text(45,+4)[lb]{$W$}
\ArrowLine(60,0)(75,15)
\Text(80,15)[lb]{$\mu$}
\Line(60,0)(75,-15)
\Text(80,-15)[lt]{$\ol \nu$}
\Photon(30,0)(30,-30){2}{3}
\Text(35,-25)[lb]{$\gamma Z$}
\ArrowLine(30,-30)(45,-45)
\Text(50,-45)[lt]{$b$}
\Line(30,-30)(15,-45)
\Text(10,-45)[rt]{$\ol b$}

\end{picture}

\caption[]{Diagrams for $e^+e^- \ra u\ \bar d\  \mu\ \bar\nu_\mu\
  b\ \bar b$}
\label{f1}
\end{figure}
\vfill


\begin{thebibliography}{1}

\bibitem{ha1} M.Jacob, J.C. Wick, \ap 7 1959 404;\\
 J.D. Bjorken, M.C. Chen, \pr 154 1966 1335;\\ 
O.~Reading-Henry \pr 154 1967 1534;


\bibitem{ha2}
P.~De~Causmaecker, R.~Gastmans, W.~Troosts, T.~T.~Wu,
pl B105 1981 215, \np B206 1982 53;\\
F.A.~Berends, R.~Kleiss,P.~De~Causmaecker, R.~Gastmans, W.~Troosts
, T.~T.~Wu,
\np B206 1982 61, B239 (1984) 382,  B239 (1984) 395,
 B264 (1986) 243,  B264 (1986) 265;\\
M. Caffo, E. Remiddi, \hpa 55 1982 339;\\
G.R.~Farrar, F.~Neri, \pl B130 1983 109;\\
G. Passarino, \pr D28 1983 2867, \np B237 1984 249\\
 F.A.~Berends, P.H.~Daverveldt, R.~Kleiss \np B253   1985  441;\\
 R.~Kleiss, W.J.~Stirling,
\np B262 1985 235,  \pl B179  1986 159;
 J.~Gunion, Z. Kunszt, \pl B161 1985 333;\\
 K.~Hagiwara, D.~Zeppenfeld, \np B274 1986 1;\\
Z.~Xu, Da-Hua Zhang, L. Chang \np B291 1987 392.

\bibitem{method} A. Ballestrero, E. Maina, \pl B350 1995 225.

\bibitem{fort}
T.~Sj\"ostrand, \cpc 82 1994 74;\\
G. J. van Oldenborgh, P. J. Franzini, A. Borrelli,\\
 \cpc  83 1994 14;\\
F.A. Berends, R. Pittau, R. Kleiss, \cpc  85 1995 437;\\
M. Skrzypek, S. Jadach, W. Placzek, Z. Was, prep. CERN-TH-95-205 (1995);\\
F. Caravaglios, M. Moretti, \pl B358 1995 332;\\
G. Passarino, \cpc 97 1996 261;\\
H. Anlauf, P. Manakos, T. Ohl, H. D. Dahmen, prep. IKDA-96-15 (1996),
hep-ph/9605457;\\
D. Bardin et al., prep.  DESY-96-233 (1996),  hep-ph/9612409;\\
D. G. Charlton, G. Montagna, O. Nicrosini, F. Piccinini, \cpc  99 1997 355;\\
C. G. Papadopoulos, \cpc 101 1997 183;\\
J. Fujimoto et al., \cpc 100 1997 128;\\
P.A. Baikov et al., prep.  HEPPH-9701412 (1997),  hep-ph/9701412.

\bibitem{cpc} E. Accomando, A. Ballestrero, \cpc 99 1997 270.

\bibitem{yr}  Physics at LEP2, G. Altarelli T.  Sjostrand, F. Zwirner eds., 
  CERN 96-01

\bibitem{wweg} D. Bardin, R. Kleiss et al.,
 {\it Event Generators for WW Physics} in ref.~\cite{yr}

\bibitem{dpeg} M. L. Mangano, G. Ridolfi et al.,
 {\it Event Generators for Discovery Physics}  in ref.~\cite{yr}


\bibitem{ww}
F.A. Berends, P.H. Daverveldt, R. Kleiss, Nucl.Phys. B253(1985)441;\\
D. Bardin, M. Bilenky, A.~Olchevski, T.~Riemann,\\
Phys. Lett. B308(1993)403; E:[ibid.B357(1995)725];\\
T. Ishikawa, T. Kaneko, S. Kawabata, Y. Kurihara, Y. Shimizu, H. Tanaka,\\
prep. KEK-92-210 (1993);\\
F.A. Berends, R. Kleiss, R. Pittau,  Nucl.Phys. B424(1994)308;\\
 Nucl.Phys. B426(1994)344; Nucl.Phys. B, Proc. Suppl. 37B(1994)163;\\
Y.~Kurihara, D.~Perret-Gallix, Y.~Shimizu, Phys. Lett. B349(1995)367;\\
G.~Montagna, O.~Nicrosini, G.~Passarino, F.~Piccinini,
 Phys. Lett. B348(1995)178;\\
C.~G.~Papadopoulos, Phys. Lett. B352(1995)144;\\
D.~Bardin,  T.~Riemann, Nucl. Phys. B462(1996)3;\\
F. Caravaglios, M. Moretti Oxford prep. OUTP-96-13-P, Apr 1996;\\
 E. Accomando, A. Ballestrero, G. Passarino, 
\np B476 1996 3.


\bibitem{hig}
E. Boos, M. Sachwitz, H.J. Schreiber, S. Shichanin, \zp C61 1994 675;\\
\zp C64 1994 391; Int. J. Mod. Phys. A10 1995 2067;\\
\zp C67 1995 613;\\
M. Dubinin, V. Edneral, Y. Kurihara, Y. Shimizu, \pl  B329 1994 379;\\
D. Bardin, A. Leike, T. Riemann, \pl B344 1995 383;\\
\pl B353 1995 513;\\
D. Apostolakis, P. Ditsas, S. Katsanevas, prep. CRETE-96-12 (1996),
hep-ph/9603383;\\
G. Montagna, O. Nicrosini, F.Piccinini, \pl B348 1995 496;\\
G. Passarino, \np B488 1997 3.


\bibitem{bbww} A. Ballestrero, E. Maina, S. Moretti, \pl B333 1994 434 \\
A. Ballestrero, E. Maina, S. Moretti, \pl B335 1994 460.

\bibitem{pv} G. Montagna, M. Moretti, O. Nicrosini, F. Piccinini
Pavia Prep. FNT/T-97/10, hep-ph/9705333

\bibitem{kuri} Y. Kurihara and F. Yuasa, private communication.

\bibitem{class}
D.~Bardin, M.~Bi\-len\-ky, D.~Leh\-ner, A.~Ol\-chev\-ski,  T.~Riemann,
Nucl. Phys. (Proc. Suppl.)  37B (1994) 148;

\bibitem{top} F. Abe et al. (CDF Coll.), \pr D50 1994 2966 \ and \prl 74 1995
2626; S. Abachi et al (D0 Coll.), \prl 74 1995 2632.

\bibitem{kuhn}
V.S. Fadin, V.A. Khoze, JETP Lett. 46 (1987) 525, 
Sov. J. Nucl. Phys. 48 (1988) 309;\\
M.J. Strassler,  M.E. Peskin, \pr D43 1991 1500;\\
M. Jezabek, J.H. K\"uhn,  T. Teubner, \zp C56 1992 653;\\
Y. Sumino, K. Fujii, K. Hagiwara, M. Murayama,  C.K. Ng, \pr D47 (1992) 56;\\
M. Jezabek,  J.H. K\"uhn \pl B316 (1993) 360.

\bibitem{ttform}
W. Bernreuther, O. Nachtmann, P. Overmann,  T. Schro\"der, \np B388 1992 53,
B406 (1993) 516;\\
G. L. Kane, G. A. Ladinsky, C. P. Yuan, \pr D45 1991 124.
\bibitem{wwz}
J. Kalinowsky \app B23 1992 1237;\\
G. B\'langer,  F. Boudjema, \pl B288 1992 201;\\
F. Boudjema, LAPP prep.  ENSLAPP-A-431-93 (1993), hep-ph/9308343.

\bibitem{miya}
A. Miyamoto, Kek prep 95-185 (1995).

\bibitem{doba}
 A. Dobado, M.J. Herrero, J.R. Pelaez, E. Ruiz Morales,  M.T. Urdiales,
\pl B352 1995 400;\\
 A. Dobado, M.T. Urdiales,  \zp C71 1996 659.

\bibitem{phact} A. Ballestrero, PHACT 1.0 - Program for Helicity
Amplitudes Calculations with Tau matrices; Torino prep. in preparation.

\bibitem{vegas} G.P.~Lepage,
{\it Jour. Comp. Phys.} { 27} (1978) 192.

\bibitem{sf} E.~A.~Kuraev,  V.~S.~Fadin, Sov.~J.~Nucl.~Phys. {\bf 41}
 (1985) 466;\\
G.~Altarelli,  G.~Martinelli, in {\it Physics at LEP}, CERN Report 86-02,
J.~Ellis, R.~Peccei eds. (CERN, Geneva 1986), vol I, pg. 47;\\
O. Nicrosini, L. Trentadue, \pl B196 1987 551.

\bibitem{circe}
T. Ohl, prep. IKDA-96-13 (1996), hep-ph/9607454.

\bibitem{twidth} M. Jezabek,  J.H. K\"uhn, \np B314 1989 1, \pr D48 1993 1910.

\bibitem{Zttcor} J. Jersak, E. Laermann,  P.M. Zerwas, {\it Phys. Rev.}
  D25 (1982) 363; J. Schwinger, 'Particles, Sources and Fields', 
 Addison-Wesley (1973); L. Reinders, H. Rubinstein,  S. Yazaki, {\it Phys. 
 Reports}  C127 (1985) 1.

\end{thebibliography}
\end{document}